\documentclass[sn-mathphys,Numbered]{sn-jnl}

\usepackage{graphicx}%
\usepackage{multirow}%
\usepackage{amsmath,amssymb,amsfonts}%
\usepackage{amsthm}%
\usepackage{mathrsfs}%
\usepackage[title]{appendix}%
\usepackage{xcolor}%
\usepackage{textcomp}%
\usepackage{manyfoot}%
\usepackage{booktabs}%
\usepackage{algorithm}%
\usepackage{algorithmicx}%
\usepackage{algpseudocode}%
\usepackage{listings}%
\usepackage{bm}
\usepackage{siunitx}
\usepackage{framed}
\usepackage{braket}
\usepackage{soul}

\theoremstyle{thmstyleone}%
\theoremstyle{thmstyletwo}%

\theoremstyle{thmstylethree}%

\begin{document}

\title[Article Title]{Half Quantum Mirror Hall Effect}


\author[1]{\fnm{Fu} \sur{Bo}}

\author[2]{\fnm{Bai} \sur{Kai-Zhi}}

\author*[2,3]{\fnm{Shen} \sur{Shun-Qing}}\email{sshen@hku.hk}

\affil[1]{\orgdiv{School of Sciences}, \orgname{Great Bay University}, \orgaddress{\city{Dongguan}, \postcode{523000}, \state{Guangdong Province}, \country{China}}}

\affil[2]{\orgdiv{Department of Physics}, \orgname{The University of Hong Kong}, \orgaddress{\street{Pokfulam Road}, \city{Hong Kong},   \country{China}}}

\affil[3]{\orgname{Quantum Science Center of Guangdong-Hong Kong-Macau Greater Bay Area},   \country{China}}


\abstract{We report the discovery of the half-quantized mirror Hall effect,
a novel quantum-anomaly induced by mirror symmetry in a strong topological
insulator (TI) film. These films are known to host a pair of gapless
Dirac cones associated with surface electrons. Our findings reveal
that mirror symmetry assigns a unique mirror parity to each Dirac
cone, resulting in a half-quantized Hall conductance of $\pm\frac{e^{2}}{2h}$
for each cone. Despite the total electric Hall conductance being null
due to time-reversal invariance, the difference in the Hall conductance
between the two cones yields a quantized Hall conductance of $\frac{e^{2}}{h}$
for the difference in mirror currents. The effect of helical edge
mirror current, a crucial feature of this quantum effect, can be 
determined by means of electrical measurements. Overall, the half-quantum
mirror Hall effect reveals a new type of mirror-symmetry induced quantum
anomaly in a time-reversal invariant lattice system, giving rise to
a topological metallic state of matter with time-reversal invariance.}




\maketitle

\section{Introduction}\label{sec1}

In the quantum field theory, the coupling of a single flavor of 2+1d massless
Dirac fermions to a U(1) gauge field results in a topological Chern-Simons
term for the gauge field, which corresponds to a half-quantized Hall
conductance. This phenomenon explicitly violates parity and time reversal
symmetries, leading to the emergence of parity anomaly \citep{Niemi1983prl,Redlich1984prl}.
Due to this anomaly, a single Dirac cone theory cannot have an ultraviolet
completion without breaking parity symmetry. In lattice crystals,
the bandwidth of the band structure is finite, and the lattice spacing
provides a natural ultraviolet cutoff for the wave-vector. As a result,
massless Dirac cones always appear in pairs in lattice systems with
time-reversal symmetry to avoid quantum anomaly. For instance, graphene
exhibits a pair of massless Dirac cones \citep{neto2009electronic},
in addition to the double degeneracy of electron spin, while a strong
topological insulator (TI) film hosts a pair of surface massless Dirac
fermions \citep{hasan2010colloquium,qi2011topological,Shen-book-2nd}.

The quest to realize parity anomaly in condensed matter has been ongoing
since the early 1980s \citep{Fradkin1986prl,Haldane1988prl}. The
primary approach involves introducing a symmetry-breaking term to
open an energy gap in the Dirac fermions \citep{Haldane1988prl,qi2008topological,yu2010quantized,qiao2010quantum,chu2011surface,li2019intrinsic}.
For example, Haldane \citep{Haldane1988prl} proposed a periodic alternating
magnetic flux in a graphene lattice to enforce gap opening in paired
Dirac cones, while Yu et al. \citep{yu2010quantized} suggested doping
transition metal elements into a magnetically ordered TI film. These
predictions have led to the observation of the quantum anomalous Hall
effect \citep{chang2013experimental,checkelsky2014trajectory,kou2014scale,deng2020quantum,liu2020robust},
with numerous efforts continuing to explore quantum anomaly in condensed
matter \citep{zhang2017anomalous,bottcher2019survival,fang2019new,mogi2022experimental,wang2021helical,fu2022quantum}. Another approach is the realization of a single gapless Dirac cone on a lattice, which breaks the time reversal symmetry to avoid the fermion doubling, and gives rise to a half quantized Hall conductance \citep{fu2022quantum}. Recent measurement of half-quantized Hall effect in a semimagnetic topological insulator \citep{mogi2022experimental} revealed the signature of parity anomaly of a single Dirac cone in a parity anomalous semimetal \citep{zou2022half,zou2023half}. The chiral edge current of a parity anomalous semimetal has recently been reported in an electrical circuits experiment\cite{yang2023realization}.

A key question is whether it is possible to directly observe parity
anomaly in a time-reversal symmetric system without introducing symmetry-breaking
perturbations, despite the constraint of symmetry. The surface states
of a topological insulator are considered a promising platform for
realizing parity anomaly in condensed matter. A topological insulator
film must have two surfaces, with anomalous terms from each surface
alternating in sign and canceling each other. The anomaly can thus
be viewed as an obstruction to regularizing a continuum theory on
a lattice without breaking symmetry or gauge invariance. In this study,
we report the discovery of a novel quantum anomaly in a mirror-symmetric
TI film with time-reversal invariance. In addition to time-reversal
symmetry, certain strong TIs, such as $\mathrm{Bi}_{2}\mathrm{Te}_{3}$
and $\mathrm{Bi}_{2}\mathrm{Se}_{3}$, exhibit additional mirror planes
perpendicular to specific axes. It is feasible to grow or to fold
mechanically a thin film with a twin boundary that respects both mirror
and time-reversal symmetry. Consequently, a mirror parity can be assigned
to the two independent Dirac cones, leading to a quantum anomaly for
each cone with a half-quantized Hall conductance of $\pm\frac{e^{2}}{2h}$.
Although the total electric Hall conductance is zero due to time-reversal
invariance, the difference in the Hall conductance between the two cones
results in a quantized Hall conductance $\frac{e^{2}}{h}$ for the
difference in mirror currents. This phenomenon is termed the half
quantum mirror Hall effect. The helical edge mirror current, a key
feature of this quantum effect, can be measured using two-terminal
transport measurements by full electric means.

\section{Results}\label{sec2}

\paragraph*{The mirror symmetry and single gapless Dirac cones-}

The surface of a $\mathrm{Z}_{2}$ strong topological insulator hosts
an  odd number of gapless Dirac surface cone as a consequence
of the bulk-surface correspondence \citep{fu2007topological,hasan2010colloquium,qi2011topological}.
For simplicity, we just consider the case of one gapless Dirac surface
cone. In the case, a TI film hosts a pair of gapless surface states
separated spatially if we assume the film is thick enough such that
the finite size effect can be ignored \citep{lu2010massive}. Denote
the Hamiltonian $\mathcal{H}$ for the quasi two-dimensional system,
which can be viewed as a semimetal with its degenerated low-energy
spectra consisting of doubled Dirac fermions. As the system does not
break the time reversal symmetry, the appearance of the double paired
Dirac cones does not causes any quantum anomaly in general if the
two bands away from the Dirac point or at higher energy part are inseparable.

The presence of the mirror symmetry in the $z$ direction $\mathcal{M}_{z}$
(perpendicular to the film as shown in left panel of Fig. 1) leads to a further classification of the
band structure \citep{hori2003mirror,teo2008surface,fu2011topological,hsieh2012topological,chiu2014classification,ando2015topological}.
$\mathcal{M}_{z}$ is an inversion operator under the sign flip of
the Cartesian coordinate component perpendicular to the mirror plane,
$(r_{x},r_{y},r_{z})\to(r_{x},r_{y},-r_{z})$. Mirror symmetry also applies
a 180 degree rotation about the z-axis to the electron spin, so $\hat{\psi}_{\uparrow}\to-i\hat{\psi}_{\uparrow}$
and $\hat{\psi}_{\downarrow}\to i\hat{\psi}_{\downarrow}$ such that the operator
squares to $-1$ for spin one half fermions, $\hat{\mathcal{M}}_{z}^{2}=-1$.
The mirror symmetry means $[\mathcal{\hat{H}},\hat{\mathcal{M}}_{z}]=0$, which
implies that all the states of the quasi-2D system can be labeled
with a mirror eigenvalue $\pm i$ as shown in Fig. 1 (middle panel). Thus, 
we can utilize the mirror symmetry to define a projection operator denoted
as $\hat{\mathcal{P}}_{\chi}$, with the expression $\hat{\mathcal{P}}_{\chi}=\frac{1}{2}(1-i\chi\hat{\mathcal{M}}_{z})$
where $\chi$ represents the eigenvalue of the mirror symmetry operation
$\hat{\mathcal{M}}_{z}$. By
utilizing the projection operator, the Hamiltonian $\hat{\mathcal{H}}$
can be decomposed into the two distinct sectors,
$\hat{\mathcal{H}}_{+}$ and $\hat{\mathcal{H}}_{-}$ where each sector
$\hat{\mathcal{H}}_{\chi}$ is given by $\hat{\mathcal{H}}_{\chi}=\hat{\mathcal{P}}_{\chi}\hat{\mathcal{H}}\hat{\mathcal{P}}_{\chi}$.
 The time reversal symmetry $\hat{\mathcal{T}}$ commutes
with any spatial symmetry $[\hat{\mathcal{T}},\hat{\mathcal{M}}_{z}]=0$, thus
the two sectors are time-reversal counterpart of each other $\hat{\mathcal{T}}\hat{\mathcal{H}}_{\chi}\hat{\mathcal{T}}^{-1}=\hat{\mathcal{H}}_{-\chi}$,
but each $\hat{\mathcal{H}}_{\chi}$ breaks the time-reversal symmetry.
Since there exist a pair of the gapless Dirac fermions within the
bulk gap of a strong TI, we come to draw a conclusion that \textit{for
a mirror-symmetric TI film, each sector $\hat{\mathcal{H}}_{\chi}$ hosts
a single gapless Dirac cone in the first Brillouin zone which is the spatial mixture of the top and bottom surface states with even or odd mirror parity.}

Existence of a single massless Dirac cone in each $\hat{\mathcal{H}}_{\chi}$
means to have parity anomaly. The half-quantized Hall conductance
is associated to each mirror sector $\sigma_{xy}^{\chi}=\frac{\chi}{2}\frac{e^{2}}{h}$
as a consequence of quantum anomaly of massless Dirac fermions \citep{fu2022quantum,zou2023half}.
The time reversal symmetry requires the co-existence of two flavors
of massless Dirac fermions with $\chi=+1$ and $-1$, and the total
Hall conductance $\sigma_{xy}^{H}=\sum_{\chi}\sigma_{xy}^{\chi}=0$.
However, the their difference defines a nontrivial mirror Hall conductance,
$\sigma_{xy}^{M_{z}}=\sum_{\chi}\chi\sigma_{xy}^{\chi}=\frac{e^{2}}{h}$.
The mirror Hall effect is quite similar to the quantum spin Hall effect
\citep{kane2005quantum,bernevig2006quantum}, but the mirror Hall
conductance is only one half of quantum spin Hall effect. Hence we
term as ``half quantum mirror Hall effect''. It is a metallic or
semi-metallic phase as the Fermi level always crosses the conduction
or valence bands of the massless Dirac fermions, which is essentially
distinct from the quantum spin Hall effect. From the band theory in
solid, the factor $\frac{\chi}{2}$ is closely associated to the band
structure of the massless Dirac cone \citep{fu2022quantum}. Thus,
the mirror symmetry induced quantum anomaly also defines a novel type
of quantum anomalous semimetal with time reversal invariance.


\paragraph*{Mirror plane with time-reversal symmetry breaking-}

The time reversal symmetry breaking in the sector $\hat{\mathcal{H}}_{\chi}$
is attributed to the existence of the mirror plane in a strong topological insulator system, which can be regarded
as an origin of quantum anomaly. The band structures of $\mathrm{Bi}_{2}\mathrm{Se}_{3}$
and $\mathrm{Bi}_{2}\mathrm{Te}_{3}$ have been investigated extensively.
The tight-binding model was first proposed to describe the topological
nature of $\mathrm{Bi}_{2}\mathrm{Se}_{3}$ and $\mathrm{Bi}_{2}\mathrm{Te}_{3}$
for the electrons of $\mathrm{P}_{z,\uparrow}$ and $\mathrm{P}_{z,\downarrow}$
orbitals from $\mathrm{Bi}$ and $\mathrm{Te}$ or Se atoms near
the Fermi energy \citep{zhang2009topological,liu2010model,chu2011surface}.
Consider the Hamiltonian of a topological insulator thin film stacked
along $z$ direction in the real-space representation $(\mathbf{r},z)$
with $\mathbf{r}=(x,y)$,
\begin{align}
\hat{\mathcal{H}} =\sum_{\mathbf{r},z}\Bigg[\hat{\psi}_{\mathbf{r},z}^{\dagger}H_{0}(z)\hat{\psi}_{\mathbf{r},z}+\Big(\sum_{\boldsymbol{\delta}=\hat{x},\hat{y}}\hat{\psi}_{\mathbf{r},z}^{\dagger}H_{\boldsymbol{\delta}}(z)\hat{\psi}_{\mathbf{r}+\boldsymbol{\delta},z}+\hat{\psi}_{\mathbf{r},z}^{\dagger}T_{z,z+1}\hat{\psi}_{\mathbf{r},z+1}+h.c.\Big)\Bigg]\label{eq:Hamiltonian}
\end{align}

where the creation field operator $\hat{\psi}_{\mathbf{r},z}^{\dagger}=(\hat{c}_{\mathbf{r},z,P1_{z}^{+},\uparrow}^{\dagger},\hat{c}_{\mathbf{r},z,P2_{z}^{-},\uparrow}^{\dagger},\hat{c}_{\mathbf{r},z,P1_{z}^{+},\downarrow}^{\dagger},\hat{c}_{\mathbf{r},z,P2_{z}^{-}\downarrow}^{\dagger})$
with the internal degrees of freedom in the unit cell including the
orbit and spin, $H_{0}(z)$ denotes the on-site Hamiltonian matrix
and $H_{\boldsymbol{\delta}}(z)$ represents the hopping matrix that
characterizes electron transitions between adjacent sites along $x$
or $y$ direction, corresponding to the displacement vectors $\boldsymbol{\delta}=\hat{\mathbf{x}}$
or $\hat{\mathbf{y}}$, respectively. $T_{z,z+1}$ and $T_{z,z+1}^{\dagger}$
correspond to the hopping terms between adjacent layers in the z-direction.  For a thin film with even number of layers $L_{z}$, as the mirror
plane located at center of the middle two layers, the indices $z$
are sequentially labeled as $-L_{z}/2,...,,-1,1,...,L_{z}/2$ as illustrated in Fig. 2a. The mirror (or reflection) symmetry $\mathcal{M}_{z}$
along $z$-axis transforms the creation and the annihilation operators
as 
\begin{align}
\mathcal{\hat{\mathcal{M}}}_{z}:\hat{\psi}_{\mathbf{r},z}^{\dagger}\to & \hat{\psi}_{\mathbf{r},-z}^{\dagger}U_{M_{z}}^{\dagger},\label{eq:symmetry_creation_operators}\\
\hat{\mathcal{M}}_{z}:\hat{\psi}_{\mathbf{r},z}\to & U_{M_{z}}\hat{\psi}_{\mathbf{r},-z}.\label{eq:symmetry_annihilation_operators}
\end{align}
Here $U_{M_{z}}=-i\sigma_{z}\tau_{z}$ is a unitary matrix representation
of $\mathcal{M}_{z}$ on the space of one site. The parity $\tau_{z}$ takes values $\pm1$ for two
orbits and $\sigma_{z}=+1$(-1) for spin up (down) in $z$ direction. If the system exhibits the mirror symmetry along z-axis, such that
$\hat{\mathcal{M}}_{z}^{-1}\hat{\mathcal{H}}\hat{\mathcal{M}}_{z}=\hat{\mathcal{H}}$,
then the following constraints hold:
\begin{align}
U_{M_{z}}^{\dagger}H_{0/\boldsymbol{\delta}}(z)U_{M_{z}} & =H_{0/\boldsymbol{\delta}}(-z),\nonumber \\
U_{M_{z}}^{\dagger}T_{z,z+1}U_{M_{z}} & =T_{-z-1,-z}^{\dagger}.\label{eq:symmetry_constraints}
\end{align}
Owing to the translational invariance in $x-y$ plane, we can apply
the Fourier transformation $\hat{\psi}_{\mathbf{k},z}=\frac{1}{\sqrt{L_{x}L_{y}}}\sum_{\mathbf{r}}e^{-i\mathbf{k}\cdot\mathbf{r}}\hat{\psi}_{\mathbf{\mathbf{r}},z}$
where $L_{x}\times L_{y}$ is the number of sites for each layer.
Subsequently, by introducing $\hat{\Psi}_{\mathbf{k}}^{\dagger}=[\hat{\psi}_{\mathbf{k},-\frac{L_{z}}{2}}^{\dagger},\hat{\psi}_{\mathbf{k},-\frac{L_{z}}{2}+1}^{\dagger},...,\hat{\psi}_{\mathbf{k},\frac{L_{z}}{2}-1}^{\dagger},\hat{\psi}_{\mathbf{k},\frac{L_{z}}{2}}^{\dagger}]$,
the Hamiltonian is expressed as $\hat{\mathcal{H}}=\sum_{\mathbf{k}}\hat{\Psi}_{\mathbf{k}}^{\dagger}H(\mathbf{k})\hat{\Psi}_{\mathbf{k}}$ where $H(\mathbf{k})$ is a  block tridiagonal matrix with respect to the layer index. In this framework, the mirror
symmetry operator is represneted by $\mathcal{M}_{z}=U_{M_{z}}\otimes I_{A}$,
where $I_{A}$ the anti-diagonal identity matrix, whose dimension
corresponds to the number of layers. It can be straightforwardly verified
that $\mathcal{M}_{z}^{-1}H(\mathbf{k})\mathcal{M}_{z}=H(\mathbf{k})$
by utilizing relations in Eq. (\ref{eq:symmetry_constraints}). Since $[H(\mathbf{k}),\mathcal{M}_{z}]=0$
,  $\mathcal{M}_{z}$ and $H(\mathbf{k})$ can be diagonalized
simultaneously. Consequently, the degenerate energy eigenstates of
$H(\mathbf{k})$ for each wave vector $\mathbf{k}$ can be distinguished
by their eigenvalues of the the mirror operator
$\mathcal{M}_{z}$, which are $i\chi$ with $\chi=\pm1$.

To more clearly comprehend the physical origins of the symmetry-breaking term within each mirror sector, we can construct an eigen-basis of mirror operator from Eqs. (\ref{eq:symmetry_creation_operators}) and (\ref{eq:symmetry_annihilation_operators})
\begin{align}
\hat{\phi}_{\mathbf{r},z,\chi}^{\dagger} & =\frac{1}{\sqrt{2}}(\hat{\psi}_{\mathbf{r},z}^{\dagger}+i\chi\hat{\psi}_{\mathbf{r},-z}^{\dagger}U_{M_{z}}^{\dagger}).\label{eq:symmetry_eigen-basis}
\end{align}

Considering that the mirror symmetry relates $z$ to $-z$ in a nonlocal manner, the two bases are constructed by employing symmetry-related pairs of unit cells located at $z$ and $-z$. In the eigen-basis, the Hamiltonian $\hat{\mathcal{H}}$
can be divided into two sectors, such that $\hat{\mathcal{H}}\equiv\sum_{\chi}\hat{\mathcal{H}}_{\chi}$, with each sector
acting on states for which the operator $\mathcal{M}_{z}$  has eigenvalues of $\pm i$. The Hamiltonian for each $\chi$ sector reads as $\hat{\mathcal{H}}_{\chi}=\hat{\mathcal{H}}_{\chi}^{0}+\hat{\mathcal{V}}_{\chi}$,
the time-reversal invariant part $\hat{\mathcal{H}}_{\chi}^{0}$ is identical
to one half of $\hat{\mathcal{H}}$ in which all the field operators $\{\hat{\psi}_{\mathbf{r},z}^{\dagger},\hat{\psi}_{\mathbf{r},z}\}$
are replaced by the eigen basis $\{\hat{\phi}_{\mathbf{r},z,\chi}^{\dagger},\hat{\phi}_{\mathbf{r},z,\chi}\}$
with all $z\ge1$ to avoid the double counting. The extra term is $\hat{\mathcal{V}}_{\chi}=-i\chi\sum_{\mathbf{r}}\hat{\phi}_{\mathbf{r},1,\chi}^{\dagger}T_{-1,1}U_{M_{z}}\hat{\phi}_{\mathbf{r},1,\chi}$ where $T_{-1,1}=t_{\perp}\sigma_{0}\tau_{z}-i\frac{\lambda_{\perp}}{2}\sigma_{z}\tau_{x}$
represents the hopping matrix connecting the neighboring layers $z=1$
and $z=-1$ around the mirror plane. 
To be more precise, $\hat{\mathcal{V}}_{\chi}$
has the form 
\begin{equation}
\hat{\mathcal{V}}_{\chi}=-\chi\sum_{\mathbf{r}}\hat{\phi}_{\mathbf{r},1,\chi}^{\dagger}(t_{\perp}\sigma_{z}\tau_{0}-\frac{\lambda_{\perp}}{2}\sigma_{0}\tau_{y})\hat{\phi}_{\mathbf{r},1,\chi}.\label{eq:symmetry_breaking_term}
\end{equation}
The first term is equivalent to a Zeeman field $-t_{\perp}$ along
the z direction and the second term represents an antiferromagnetic
order which is staggered along $z-$ direction and uniform within
$xy$ plane \citep{li2010dynamical,sekine2014axionic}, both breaking
time reversal symmetry explicitly. An intuitive explanation for the emergence of the symmetry-breaking in the mirror eigenbasis can be described as follows: The spin behaves as
a pseudo-vector or axial vector, which means that under mirror symmetry,
its component parallel to the mirror plane is inverted, while the
component perpendicular to the plane remains unchanged. From Eq. (\ref{eq:symmetry_eigen-basis}), two in-plane spins with opposite orientations located at positions symmetric with respect to a mirror plane can be arranged to maintain mirror symmetry. However, the existence of a horizontal mirror symmetry $\mathcal{M}_{z}$ requires that the spin polarization on self-reflected surfaces be oriented out-of-plane within each mirror sector. The symmetry broken term $\hat{\mathcal{V}}_{\chi}$ makes it possible that there
exists a single Dirac cone in the first Brillouin zone in $\hat{\mathcal{H}}_{\chi}$
on a lattice [see Fig. 1(right panel)].   As shown in Fig. 2b, we plot the energy spectrum of the topological insulator thin film  calculated using the tight-binding model [Eq.(\ref{eq:Hamiltonian})] with a thickness of $L_z=80$, shown in  gray solid lines. Alongside, we show the spectrum of the Hamiltonian $\hat{\mathcal{H}}_{\chi}$  in the mirror-eigenbasis (red circles), which considers a halved thickness of $L_z/2=40$ and extra symmetry breaking term [Eq. (\ref{eq:symmetry_breaking_term})] on one surface. Remarkably, the two spectra coincide. Since the spectra of $\hat{\mathcal{H}}_{\chi=+}$ and $\hat{\mathcal{H}}_{\chi=-}$ are degenerate, we present only the results for $\hat{\mathcal{H}}_{\chi=+}$ in Fig. 2b.

\paragraph*{Gapless Dirac Cones with Parity Symmetry Breaking-}

We then develop a gapless Dirac cone 
model for mirror-symmetric topological insulator thin film to elucidate
the fundamental physics underlying half quantum mirror Hall effect.
Our focus is on the geometry depicted in Fig. 2a, which is characterized
by an open boundary condition along the z-axis and periodic boundary
conditions in the xy-plane. The system's Hamiltonian can be decomposed
into two seperate parts: $H(\mathbf{k})=H_{1d}(\mathbf{k})+H_{\parallel}(\mathbf{k})$, both exhibiting the mirror symmetry. $H_{1d}(\mathbf{k})$ is a one-dimensional lattice model for topological insulator film with a momentum-dependent gap $M_0(\mathbf{k})+2t_{\perp}$. $H_{\parallel}(\mathbf{k})$ is  block-diagonalized within the layer space.
We first address the eigenproblem of $H_{1d}(\mathbf{k})$, where
the eigenvalue equation is given by $H_{1d}(\mathbf{k})|\Phi_{n,\zeta,\chi,\mathbf{k}}\rangle=\zeta \Delta_{n}(\mathbf{k})|\Phi_{n,\zeta,\chi,\mathbf{k}}\rangle$
with $|\Phi_{n,\zeta,\chi,\mathbf{k}}\rangle$ being the eigenvectors and $\zeta \Delta_{n}(\mathbf{k})$ 
the corresponding eigenvalues with $\zeta=\pm$ and $n=1,2,3,...,L_z$. $i\chi=\pm i$ represent the mirror
eigenvalues of the degenerate energy eigenstates of $H_{1d}(\mathbf{k})$
for each wave vector $\mathbf{k}$, i.e. $\mathcal{M}_{z}|\Phi_{n,\zeta,\chi,\mathbf{k}}\rangle=i\chi|\Phi_{n,\zeta,\chi,\mathbf{k}}\rangle$.
Upon obtaining the eigenstates of $H_{1d}(\mathbf{k})$, we proceed
to project the remaining part of the Hamiltonian $H_{\parallel}(\mathbf{k})$
onto this eigenbasis. In the eigenproblem, we aim to determine a solution
of the form $\Phi(z)\sim e^{i\xi z}\Phi(0)$ where $\xi$ is a general
complex number. In our model $H_{1d}(\mathbf{k})$, the $z$-component
of spin denoted as $s=\pm$ is conserved: $[H_{1d}(\mathbf{k}),\sigma_{z}]=0$. Moreover, since the spin z operator $\sigma_z$ commutes with the mirror symmetry operator $\mathcal{M}_{z}$ (i.e., $[\sigma_z,\mathcal{M}_{z}]=0$), $\sigma_z$ and
$\mathcal{M}_{z}$ can be diagonalized simultaneously. As a consequence, 
the eigenstates can be relabeled as $|\Phi_{n,\zeta,s,\mathbf{k}}\rangle=|\phi^{s}_{n,\zeta,\mathbf{k}}\rangle\otimes|s\rangle$
where $|s\rangle$ is an eigenstate of $\sigma_{z}$ with eigenvalues
$s=\pm$ , i.e. $\sigma_{z}|s\rangle=s|s\rangle$. The corresponding eigen
equation for the spatial components $\phi^{s}_{n,\zeta,\mathbf{k}}(z)$ becomes
\begin{equation}
\{[M_{0}(\mathbf{k})+2t_{\perp}\cos\xi]\tau_{z}+s\frac{\lambda_{\perp}}{2}\tau_{x}\sin\xi\}\phi^{s}_{n,\zeta,\mathbf{k}}(z)=\zeta\Delta_n(\mathbf{k})\phi^{s}_{n,\zeta,\mathbf{k}}(z).\label{eq:eigenfunction_for_single_spin}
\end{equation}
 If $\xi$ solves this equation, then $-\xi$ is also a solution.
 By substituting
general solutions that adhere to the boundary condition into the eigen
equations, we derive a set of equations that are self-contained and
can be solved to determine the eigenvalues $\zeta\Delta_n(\mathbf{k})$ and also the corresponding eigenstates $|\phi_{n,\zeta,\mathbf{k}}^{s}\rangle$. We ascertain that these eigenstates are indeed the eigenfunctions of the mirror operator: $\mathcal{M}_z |\Phi_{n,\zeta,s,\mathbf{k}}\rangle=-is\zeta|\Phi_{n,\zeta,s,\mathbf{k}}\rangle$. Given that $H_{\parallel}(\mathbf{k})$ is invariant under mirror symmetry: $[\mathcal{M}_z,H_{\parallel}(\mathbf{k})]=0$, also considering that it flips the spin, the projection onto these eigenbasis reveals that only the states with opposite $\zeta$ and opposite spin configuration have non-zero overlap: $\langle\Phi_{n^\prime,\zeta^\prime,s^\prime,\mathbf{k}}|H_{\parallel}(\mathbf{k})|\Phi_{n,\zeta,s,\mathbf{k}}\rangle=\lambda_{\parallel}\delta_{\zeta^\prime,-\zeta}\delta_{n,n^\prime}[\sin(k_x)\sigma_x+\sin(k_y)\sigma_y]_{s^\prime,s}$. Therefore, the Hamiltonian $H_{\parallel}(\mathbf{k})$ couples basis states with opposite spins within the same mirror eigensector.(See Supplementary Material for further details.)

After projection, we identify a series of Dirac bands: four are gapless, while the remaining bands are gapped and topologically trivial. The topological phenomena manifest in the four gapless bands; here, the low-energy states correspond to surface states, and the high-energy states within these bands transition to bulk states (as shown in Fig. S2). In the eigen basis of mirror symmetry,
the wave functions of the surface states in each sector $\hat{\mathcal{H}}_{\chi}$
are symmetric ($\chi=+1$) or antisymmetric ($\chi=-1$) about $z$,
in which the sites $z$ and $-z$ are connected and breaks the locality
property on the lattice. By solving the three-dimensional model for
a TI film, we find it essential to incorporate a term that breaks the symmetry to provide and accurate representation of the surface states\citep{zou2023half}:

\begin{equation}
H_{\mathrm{surf},\chi}=\lambda_{\parallel}\sin k_{x}\widetilde{\sigma}_{x}+\lambda_{\parallel}\sin k_{y}\widetilde{\sigma}_{y}+\chi \Delta(\mathbf{k})\widetilde{\sigma}_{z}\label{eq:effective_H}
\end{equation}
where the Pauli matrices $\widetilde{\boldsymbol{\sigma}}$ act on the spaces spanned by  $[|\Phi_{n=1,\zeta=+,\chi=+}\rangle,|\Phi_{n=1,\zeta=-,\chi=+}\rangle]$ and $[|\Phi_{n=1,\zeta=-,\chi=-}\rangle,|\Phi_{n=1,\zeta=+,\chi=-}\rangle]$ for the $\chi=+$ and $\chi=-$  sectors respectively and $ \Delta(\mathbf{k})=\Theta[-m_{0}(\mathbf{k})]m_{0}(\mathbf{k})$ with $\Theta(x)$ as a step function
and $m_{0}(\mathbf{k})=m_{0}-4t_{\parallel}(\sin^{2}\frac{k_{x}}{2}+\sin^{2}\frac{k_{y}}{2})$.
$m_{0}$ is the bulk gap of TI. Near the $\Gamma$ point, $\Delta(\mathbf{k})=0$
for $m_{0}(\mathbf{k})>0$ and $H_{\mathrm{surf},\chi}$ preserves the parity
symmetry while $\Delta(\mathbf{k})=m_{0}(\mathbf{k})$ for $m_{0}(\mathbf{k})<0$ breaks the parity
and time reversal symmetry. For the detail of the derivation, see
Supplementary Information. With the inclusion of the symmetry-breaking term, the gapless states described by the Hamiltonian in Eq. (\ref{eq:effective_H}) diverges significantly from the conventional Dirac surface states\citep{qi2011topological,Konig2008jpsj}. Under vertical mirror symmetry
(also known as parity symmetry in two dimensions), the annihilation
operators transform as $\hat{\mathcal{M}}_{x}:\hat{\psi}_{(x,y,z)}\to U_{M_{x}}\hat{\psi}_{(-x,y,z)}$
where $U_{M_{x}}=-i\sigma_{x}\tau_{z}$. From Eq.
(\ref{eq:symmetry_annihilation_operators}), we find the horizontal
and vertical mirror operators anticommute: $\{\hat{\mathcal{M}_{x}},\hat{\mathcal{M}}_{z}\}=0$.
Then for an eigenstate $|\psi_{\chi}\rangle$ of $\hat{\mathcal{M}}_{z}$
that satisfies $\hat{\mathcal{M}}_{z}|\psi_{\chi}\rangle=i\chi|\psi_{\chi}\rangle$,
it follows that $\hat{\mathcal{M}}_{z}\hat{\mathcal{M}_{x}}|\psi_{\chi}\rangle=-\hat{\mathcal{M}_{x}}\hat{\mathcal{M}}_{z}|\psi_{\chi}\rangle=-i\chi\hat{\mathcal{M}_{x}}|\psi_{\chi}\rangle$.
This implies that after the tranformation of $\hat{\mathcal{M}_{x}}$,
the state $|\psi_{\chi}\rangle$ flips its eigenvalue of $\hat{\mathcal{M}_{z}}$,
$\hat{\mathcal{M}_{x}}|\psi_{\chi}\rangle\sim|\psi_{-\chi}\rangle$.
The eigenvalue of $\hat{\mathcal{M}}_{z}$ defines the chirality with
respect to $\hat{\mathcal{M}_{x}}$, with $|\psi_{\chi=+}\rangle$
being left-handed state while $|\psi_{\chi=-}\rangle$ being the right-handed
state. As shown in Fig. 2c, we present a schematice diagram to illustrate the main difference between the proposed gapless Dirac cone model with broken parity symmetry ($\hat{\mathcal{M}_{x}}$) and the conventional Dirac surface states:  in the parity invariant regime ($m_{0}(\mathbf{k})>0$), the pesudo-spin texture is confined to the  x-y plane, whereas outside this regime, the pesudo-spin texture acquires z components that breaks the parity symmetry explicitly. The parity symmetry (or time reversal symmetry) maps one mirror sector onto the other.  We can combine the spinors associated to each Dirac point
in Eq. (\ref{eq:effective_H}) into a four-component spinor, then the Hamiltonian becomes $H_{\mathrm{surf}}=\lambda_{\parallel}\sin k_{x}\widetilde{\tau}_{0}\widetilde{\sigma}_{x}+\lambda_{\parallel}\sin k_{y}\widetilde{\tau}_{0}\widetilde{\sigma}_{y}+m(k)\widetilde{\tau}_{z}\widetilde{\sigma}_{z}$
with $\widetilde{\boldsymbol{\tau}}$ act on the mirror space. For a Hamiltonian
constructed with three anticommuting Dirac matrices, there exists
only one additional matrix ($\widetilde{\tau}_z$), apart from the identity matrix, that commutes
with the Hamiltonian. This commutation indicates the presence of a
conserved quantum number. In the case of 3D massless Dirac fermions,
this conserved quantity is known as chirality. In the context we are
considering, the analogous conserved quantity is related to mirror
symmetry $\hat{\mathcal{M}}_z$. In Fig. 2b, the spectrum of the four-band Hamiltonian [Eq.(\ref{eq:effective_H})] are plotted as  green dashed lines, demonstrating that since this model is derived from the tight-binding model, it accurately reproduces the spectrum not only at low energies but also at the corners of the Brillouin zone (as shown in Fig. S1). In this way a single Dirac cone
may exist in the first Brillouin zone as a consequence of the symmetry
breaking to avoid the fermion doubling problem \citep{nielsen1981no}.
This is distinct from the conventional effective model for the surface
states which is only valid for a small $k$ \citep{Konig2008jpsj}. As explained in the "Topological field theory for quantum mirror Hall effect on a lattice" in the Methods section, we elucidate the relationship and distinction between our lattice-based theory and the quantum anomaly originating from an effective $k\cdot p$ model.
The mass term can be interpreted as a natural regularization that emerges on a lattice and resolves the divergence in the charge-charge and mirror-mirror polarization tensors inherent in the effective $k\cdot p$ model. Additionally, it serves as the topological origin of the half-quantum mirror Hall effect. 

We note that prior research  by Creutz and Horv$\acute{\mathrm{a}}$th \citep{Creutz1994prd} has similarly examined topological systems in film geometry. However, in constrasting with our research, we note significant distinctions.  Ref. \citep{Creutz1994prd} explored 1+D (D=1,3)-dimensional film without time-reversal symmetry and featuring chiral surfaces states, while our study investigates three-dimensional topological insulator film with time reversal symmetry and characterized by helical surface states. Consequently, the models in these works fall into two distinct topological classes, leading to fundamentally different quantum anomalies: Ref. \citep{Creutz1994prd} addresses the chiral anomaly, while our work is focused on the parity anomaly. Ref. \citep{Creutz1994prd} has primarily focused on the low-energy surface states, claiming that the degeneracy between a pair of gapless Dirac fermions cancels the anomaly. In contrast,  our study classifies Dirac cones according to parity by utilizing mirror symmetry,  revealing the persistence of the parity anomaly. This method not only reveals the full energy dispersions for each class but also highlights the critical role of mirror symmetry in generating quantum anomaly phenomena within the  topological insulator thin film.

\paragraph*{Quantum mirror Hall conductance-}

The intrinsic mirror Hall conductance can be evaluated by means of
the Kubo formula in the linear response theory \citep{mahan1981many,murakami2006quantum,yang2006stvreda},
\begin{equation}
\sigma_{xy}^{M_z}=\frac{2e}{hL_{x}L_{y}}\sum_{\mathbf{k},n\ne m}f(\epsilon_{n})\frac{\mathrm{Im}\langle u_{n\mathbf{k}}|\mathbf{j}_{M_z,x}|u_{m\mathbf{k}}\rangle\langle u_{m\mathbf{k}}|v_{y}|u_{n\mathbf{k}}\rangle}{(\epsilon_{n}-\epsilon_{m})^{2}}\label{eq:mirror_hall}
\end{equation}
where $n,m$ are the band indices, the velocity operator at each $\mathbf{k}$ is
given by $v_{i}=\frac{1}{\hbar}\frac{\partial H(\mathbf{k})}{\partial k_{i}}$
with $i=x,y$ and $\mathbf{j}_{M_z}=ie\mathcal{M}_{z}\mathbf{v}$ is
the mirror current operator,$|u_{n\mathbf{k}}\rangle$ is the eigenvector of $H(\mathbf{k})$ with the eigen-energy as $\epsilon_n(\mathbf{k})$ and $f(\epsilon_n)=\Theta(\mu-\epsilon_n)$ is the
the Fermi-Dirac distribution at zero temperature with $\mu$ as the chemical potential. 
 By using the mirror operator's eigenbasis, denoted by $|u_{n\mathbf{k}}^{\chi}\rangle$ with $\mathcal{M}_z|u_{n\mathbf{k}}^{\chi}\rangle=i\chi|u_{n\mathbf{k}}^{\chi}\rangle$,
the Kubo formula for $\sigma_{xy}^{M_z}$ can be recast as $\sigma_{xy}^{M_z}=\frac{e}{h}\sum_{\chi,n,\mathbf{k}}\chi f(\epsilon_{n}^{\chi}(\mathbf{k}))\Omega_{n}^{\chi}(\mathbf{k})$
where $\Omega_{n}^{\chi}(\mathbf{k})=2\mathrm{Im}\langle\partial_{x}u_{n\mathbf{k}}^{\chi}|\partial_{y}u_{n\mathbf{k}}^{\chi}\rangle$
is the mirror-resolved Berry curvature for each state. Each mirror sector $\hat{\mathcal{H}}^{\chi}$ belongs to the class
A of topological classifications, enabling the association of an anomalous
Hall conductance $\sigma_{xy}^{\chi}$ with it,  and the mirror  Hall conductance can be expressed as $\sigma_{xy}^{M_z}=\sum_\chi \chi\sigma_{xy}^{\chi}$.    Since 
$\hat{\mathcal{H}}_{\chi}$ contains a single gapless Dirac cone in the whole Brillouin
zone, the Stoke's theorem allows the Berry curvature integral
over the occupied states to be converted into a line integral of
the Berry connection along the Fermi surface (FS) for a partially filled band $n$: $\sigma_{xy,n}^{\chi}=\frac{e^2}{2\pi h}\int d^2\mathbf{k}\Omega_n^\chi(\mathbf{k})\Theta(\mu-\epsilon_n^{\chi})=\frac{e^2}{2\pi h} \oint_{\mathrm{FS}} d\mathbf{k}\cdot\boldsymbol{\mathcal{A}}_n^{\chi}(\mathbf{k})$ where $\boldsymbol{\mathcal{A}}_n^{\chi}(\mathbf{k})=-i\langle u_{n\mathbf{k}}^{\chi}|\partial_{\mathbf{k}}|u_{n\mathbf{k}}^{\chi}\rangle$ denotes the Berry connection. If Fermi surface consists of a single gapless Dirac
cone, the Berry phase around the Fermi surface is quantized to
$\pi$. As a result, the Hall conductance is half quantized $\sigma_{xy}^{\chi}=\chi\frac{e^{2}}{2h}$
when the chemical potential is located within the bulk gap \citep{zou2023half}.The half-quantum mirror Hall effect can also be interpreted using the gapless Dirac cone model presented in Eq. (\ref{eq:effective_H}). The Hall conductance of a generic two-band Hamiltonian $H_{\mathrm{surf},\chi}(\mathbf{k}) = \mathbf{d}_{\chi}(\mathbf{k}) \cdot \boldsymbol{\sigma}$ is given by $\sigma_{xy}^{\chi} = \frac{e^{2}}{h}\frac{1}{4\pi}\int d^2\mathbf{k} [\Theta(\mu-|\mathbf{d}_{\chi}|)-\Theta(\mu+|\mathbf{d}_{\chi}|)]\hat{\mathbf{d}}_{\chi} \cdot (\partial_{k_{x}}\hat{\mathbf{d}}_{\chi} \times \partial_{k_{y}}\hat{\mathbf{d}}_{\chi})$, which represents the coverage of the unit vector $\hat{\mathbf{d}}_{\chi}(\mathbf{k}) = \mathbf{d}_{\chi}(\mathbf{k})/|\mathbf{d}_{\chi}(\mathbf{k})|$ across the Bloch sphere for the occupied states. The Hamiltonian corresponds to vectors $\mathbf{d}_{\chi}(\mathbf{k}) = (\lambda_{\parallel}\sin k_{x}, \lambda_{\parallel}\sin k_{y}, \chi\Delta(\mathbf{k}))$. At $\mathbf{k} = (\pi,\pi)$, the unit vector $\hat{\mathbf{d}}_{\chi}(\mathbf{k}) = (0,0,\chi\mathrm{sgn}(\Delta(\pi,\pi)) )$ points to the north (south) pole on the unit sphere for $\chi = +1$ ($\chi = -1$), assuming $\mathrm{sgn}(\Delta(\pi,\pi)) > 0$.  For wavevectors on the Fermi surface which is in the vicinity of the Dirac points, the unit vector $\hat{\mathbf{d}}_{\chi}(\mathbf{k}) = (\lambda_{\parallel}\sin k_{x}, \lambda_{\parallel}\sin k_{y},0)$ resides in the equatorial plane of the unit sphere. As depicted in Fig. 2d, this configuration of $\hat{\mathbf{d}}_{\chi}(\mathbf{k})$ spans half of the unit sphere, resulting in a winding number of $ \chi/2$, which corresponds to a Hall conductance of $\sigma_{xy}^{\chi} = \chi\frac{e^{2}}{2h}$. Then, the mirror Hall conductance is quantized, $\sigma_{xy}^{M_z}=\sigma_{xy}^{\chi=+}-\sigma_{xy}^{\chi=-}=\frac{e^{2}}{h}$.To further validate it, we calculate the mirror Hall conductance as a function of the chemical potential $\mu$
using Eq. (\ref{eq:mirror_hall}) based on the tight-binding model, shown as the black line with squares in Fig. 2c. The green dashed line is according to the gapless Dirac model  in Eq. (\ref{eq:effective_H}). For comparision,  we also present the the contributions from the four lowest-energy gapless bands within tight-binding model, indicated by the blue line marked with triangles. They show good agreement with  each other. Note that as the chemical potential enters the region of bulk states (i.e., $|\mu| > |m_0|$), the topologically trivial gapped bands also begin to contribute. 

\paragraph*{Transport signature-}

The half-quantum mirror Hall effect is very similar to the spin Hall
effect in the semiconductor and can be measurable by full electric
means \citep{hirsch1999spin,brune2010evidence,balakrishnan2013colossal,sinova2015spin,kondou2016fermi}.
We consider a two-terminal transport measurement and a charge current
is driven between injector and collector electrodes by an electric
field. We first examine a system
with no coupling between two mirror sectors, where each sector individually
satisfies the equation $\sum_{j}\sigma_{ij}^{\chi}E_{j}=J_{i}^{\chi}$ with $i,j=x,y$.
When an external electric field $E_{x}$ is applied in the x-direction, electrons with opposite mirror eigenvalues acquire anomalous transverse velocities in opposite directions. This results in a transverse current that causes
charge to accumulate along the lateral edges of the material, leading
to the development of an internal electric field $E_{y}^{\chi}$ in
the y-direction for each mirror sector. This electric field opposes
further accumulation of charge thereby reaching an equilibrium. At
equilibrium, the net transverse current becomes zero (i.e., $J_y^{\chi}=0$), which allows
us to determine the equilibrium self-building electric field as $E_{y}^{\chi}=-E_{x}\frac{\sigma_{yx}^{\chi}}{\sigma_{yy}^{\chi}}$.
 Electrons with opposite mirror eigenvalues accumulate on opposite edges of the material, leading to a spatially varying mirror polarization density $\delta n_{M_z}(x)$, as illustrated in Fig. 3a. Consequently, the self-building electric fields for each mirror sector are oriented in opposite directions. This self-building electric field $E_{y}^{\chi}$
then induces a Hall current in the x-direction, described by $J_{c,x}^{\mathrm{mH}}=\sum_\chi\sigma_{xy}^{\chi}E_{y}^{\chi}$. To compute the total longitudinal charge (c) current $J_{c,x}$ we sum
the conductive current from the surface states, $J_{c,x}^{\mathrm{ss}}=\sum_{\chi}\sigma_{xx}^{\chi}E_{x}$
with the inverse Hall current, yielding $J_{c,x}=E_{x}\sum_{\chi}(-\sigma_{xy}^{\chi}\frac{\sigma_{yx}^{\chi}}{\sigma_{yy}^{\chi}}+\sigma_{xx}^{\chi})$. The charge current contains
two parts $J_{c,y}=J_{c,y}^{\mathrm{ss}}+J_{c,y}^{\mathrm{mH}}$:
the first term $J_{c,y}^{\mathrm{ss}}$ comes from the conducting
surface states and the second term $J_{c,y}^{\mathrm{mH}}$
arises from the spatial accumulation of the mirror polarization density
and is an effect due to the existence of the mirror Hall effect. This
mirror Hall mediated charge transport can be understood as follows:
the electric field first induces a mirror charge accumulation on the
boundary via the mirror Hall effect and then is converted into the
charge current along the electric field via the inverse mirror Hall
effect \citep{abanin2011giant,shimazaki2015generation,yamamoto2015valley}. By making further assumption that $\sigma_{xx}^{\chi}=\sigma_{yy}^{\chi}$, the two terminal resistance measured, as depicted in upper panel of Fig. 3b, can be expressed as:
\begin{equation}
R=\frac{L}{W\sigma_{c}}\left[1+\tan^2(\theta_{M_z})\right]^{-1}\label{eq:no_inter_scattering_resistance}
\end{equation} 
where we introduce the mirror Hall angle $\theta_{M_z}=\tan^{-1}(\sigma_{xy}^{M_z}/\sigma_{c})$ with $\sigma_c=\sum_{\chi}\sigma_{xx}^{\chi}$ as the total longitudinal charge
conductivity and employ the relation $|\tan(\theta_{M_z})|=|\sigma_{xy}^{\chi}/\sigma_{xx}^{\chi}|$. Next, we will consider the effects of scattering between the two mirror sectors.  In this situation, by solving the combined equations of generalized Ohm's law and continuity equations for currents \citep{abanin2009nonlocal,beconcini2016nonlocal,song2019low,sekine2020valley},
subject to appropriate boundary conditions, the mirror polarization
density and the charge current density are found to be (see Methods)
\begin{align}
\delta n_{M_z}(x) & =-\frac{\sigma_{xy}^{M_z}E_{y}l_{M_z}}{D_{c}}\frac{\sinh(x/l_{M_z})}{\cosh(W/2l_{M_z})},\label{eq:charge_polarization}\\
J_{c,y}(x) & =\sigma_{c}E_{y}\left[1+\tan^{2}(\theta_{M_z})\frac{\cosh(x/l_{M_z})}{\cosh(W/2l_{M_z})}\right]\label{eq:2terminal_current_density}
\end{align}
for $|x|\le W/2$ and $0$ for outside the system. $l_{M_z}$ is the inter-mirror scattering length and
$D_{c}$ is the charge diffusion constant.  
We assume inter-mirror scattering time is much longer than the scattering
time within the same mirrors eigenvalues such that $l_{M_z}\gg l_{e}$
where $l_{e}$ is the mean free path. To ensure the diffusive transport
is 2D, the width is also required to be much longer than mean-free
path, i.e. $W\gg l_{e}$. Depending on the relative amplitude of $l_{M_z}$
and $W$, we have two regimes: (i) the weak inter-mirror scattering
regime $l_{M_z}\gg W\gg l_{e}$, The polarization density variation
$\delta n_{M_z}(x)\simeq-\sigma_{xy}^{M_z}E_{y}x/D_{c}$ shows a linear
behavior along the direction perpendicular to the electric field and
independent on $l_{M_z}$ and the induced charge current is uniformly
distributed in the sample which are shown by the black lines in Fig. 3c and d respectively; (ii) the strong inter-mirror scattering
regime $W\gg l_{M_z}\gg l_{e}$, the boundary effect becomes dominate
that mirror charge accumulates at the boundaries and $J_{c,y}^{\mathrm{mH}}$
only flows near the boundaries as shown by the red lines in Fig. 3c
and d respectively. As depicted in upper panel of Fig. 3b, total charge current can be obtained by integrating
the current density over the width $I_{c,y}=\int_{-W/2}^{W/2}dxJ_{c,y}(x)$
and the voltage drops over the length $L$ of the system is $V=E_{y}L$.
From Eq. (\ref{eq:2terminal_current_density}), the two-terminal measurement
resistance $R=V/I_{c,y}$ can be expressed as
\begin{equation}
R=\frac{L}{W\sigma_{c}}\left[1+\tan^2(\theta_{M_z})\frac{\tanh{(W/2l_{M_z})}}{W/2l_{M_z}}\right]^{-1}.\label{eq:resistance}
\end{equation}
By varying the chemical potential $\mu$ through a gate voltage,
$\sigma_{c}$ can be monotonically tuned to the minimal value $\sigma_{min}=\frac{2}{\pi}\frac{e^{2}}{h}$
at the Dirac point \citep{ziegler2006robust,ostrovsky2006electron}.
While the gate voltage has little impact on the half quantum mirror
Hall effect as its root lies in the quantum anomaly of gapless Dirac
bands and is contributed by the deep-lying states.  As shown in Fig.
3e, we plot $R$ as a function of $\mu$ for different $l_{M_z}/W$.
For $l_{M_z}\gg W$, the mirror eigenvalue can be viewed as a good quantum
number, Eq. (\ref{eq:resistance}) is then reduced to Eq. (\ref{eq:no_inter_scattering_resistance}) as $\frac{\tanh(W/2l_{M_z})}{W/2l_{M_z}}\sim 1$. In this case, as $\mu$ moves away from the Dirac point, $R$ initially
rises,  subsequently peaks when $\sigma_{c}=|\sigma_{xy}^{M_z}|$, and ultimately decreases, as shown by the darkest
green line. As the hybridization between two mirror sectors
becomes stronger, $l_{M_{z}}$ reduces, leading to a decreased contribution from mirror Hall effect, as described by Eq. (\ref{eq:resistance}). When $l_{M_z}/W$ approaches zero, $R$ converges to the transport behavior associated with conventional Dirac surface states in the absence of the mirror Hall effect\citep{Bolotin2008prl,Du2008NN}, as shown by the darkest red line. 
The resistivity $R_{0}=L/W\sigma_{c}$ can be measured by using conventional six-probe measurement. Thus the value of $\sigma^{M_z}_{xy}$ can be deducted from the
measurement of $R$ if the exact mirror  symmetry nearly holds (i.e., $l_{M_z}\to\infty$).

We next consider a multi-terminal measurement and the current $I$ is applied
along the y-direction, flowing from from terminal 8 to terminal 2, as depicted in the lower panel of Fig. 3b. The detailed calculations for this setup are provided in  "Methods".
Here we consider the weak inter-mirror
scattering regime $l_{M_{z}}\gg L,W$. For practical analysis, we derive the analytical
expressions in certain limiting cases: (i) In the regime where $x\gg l_{M_{z}}$, the nonlocal
resistivity associated with the mirror Hall effect 
\begin{equation}
R_{\mathrm{NL}}(x)\simeq\frac{1}{\sigma_{c}}\frac{L}{2\xi_{\mathrm{w}}}\frac{\tan^{2}(\theta_{M_{z}})}{1+\tan^{2}(\theta_{M_{z}})}e^{-x/\xi_{\mathrm{w}}}\label{eq:RNL}
\end{equation}
with $\xi_{\mathrm{w}}=l_{M_{z}}\sqrt{1+\tan^{2}(\theta_{M_{z}})}$.
(ii)Conversely, for $L,W\gg x$, we have $\omega(k)\simeq l_{M_{z}}^{-1}$
and the boundary condition along $x$ direction becomes negligible.Under these conditions,  
Eq. (\ref{eq:4resistance}) can be approximated by its value at $x=0$,
\begin{equation}
R_{28,28}\simeq\frac{L}{W\sigma_{c}}\frac{\mathcal{F}(\frac{L}{W})}{[1+\tan^{2}(\theta_{M_{z}})]}\label{eq:R2828}
\end{equation}
which is similar to the two-terminal case except for a sample sized
dependent renormalization factor $\mathcal{F}(y)=\mathrm{Re}\sum_{s=\pm}\frac{s}{\pi^{2}}\left(\text{Li}_{2}\left(e^{-s\pi/y}\right)-4\text{Li}_{2}\left(e^{-s\frac{\pi}{2y}}\right)\right)$
where $\mathrm{Li}_{2}$ is the polylogarithm function of order two. We present a plot of the resistivity as a function of the probe position $x$  based on the full integral in Eq. (\ref{eq:4resistance})  in Fig. 3f. As demonstrated, the numerical results are in good agreement with the analytical expression given in Eq. (\ref{eq:RNL}) when the voltage -measuring probes are positioned significantly far from the ternimals where the current is injected. We additionally plot the resistivity as a function of the sample aspect ration $L/W$ when voltage-measurement probes are located at the terminals where the current  the current is injected, as depicted in Fig. 3g. It is evident that when $L\ll W$, the results from the full integral are consistent well with approximation given by Eq. (\ref{eq:R2828}). Notably, as $L/W$ approaches zero, we find that $\mathcal{F}$ tends towards unity, and $R_{28,28}$ simplifies to the expression in Eq. (\ref{eq:no_inter_scattering_resistance}).
 
Finally we address the question of how to extract the half quantum mirror Hall conductivity from electrical measurements. In experiments, there are three unkown paremeters that need to be determined: the longitudinal conductivity $\sigma_c$,the mirror Hall conductivity $\sigma_{xy}^{M_z}$, and the inter-mirror scattering length $l_{M_z}$. Therefore, at least three measurements are necessary. Firstly, according to Eq. (\ref{eq:RNL}),  $\xi_{\mathrm{w}}$  can be deduced  through two nonlocal voltage measurements. These measurements are taken  between terminals 3 and 7,  and  between terminals 4 and 6,  which are at distances $\Delta x$ and $2\Delta x$ from the current injection ternimal 2 and 8, respectively, as illustrated in the lower panel of Fig. 3b.  From these measurements, $\xi_{\mathrm{w}}$ which is can be extracted by using the relation $\xi_{\mathrm{w}}=\Delta x/\ln(R_{37,28}/R_{46,28})$. Then from Eq. (\ref{eq:R2828}), after conducting a two-terminal measurement, the mirror Hall angle can be determined by the following equation:
\begin{equation}
\tan(\theta_{M_z}^{\mathrm{exp}})=\sqrt{\frac{2\Delta x}{W}\frac{R_{37,28}^2/(R_{28,28} R_{46,28})}{ \ln(R_{37,28}/R_{46,28})}}.
\end{equation}
The superscript "$\mathrm{exp}$" indicates the experiment measurement results, which are used to  distingish them from the theoretical predictions. The mirror Hall conductivity and longitudinal conductivity can be determined from the measured data using  
\begin{equation}
\sigma_{c}^{\mathrm{exp}}=\frac{1}{R_{28,28}}\frac{L}{W}\frac{1}{1+\tan^2(\theta_{M_z}^{\mathrm{exp}})}
\end{equation}
 and 
\begin{equation}
\sigma_{xy}^{M_z,\mathrm{exp}}=\frac{1}{R_{28,28}}\frac{L}{W}\frac{\tan(\theta_{M_z}^{\mathrm{exp}})}{1+\tan^2(\theta_{M_z}^{\mathrm{exp}})}\label{eq:exp_mirror_hall}.
\end{equation}
By  following this procedure, the desired electrical properties can be accuratedly determined from the measured values. We employ Eq. ($\ref{eq:4resistance}$) to assesse the validity of Eq. (\ref{eq:exp_mirror_hall}). As depicted in Fig. 3h, when the two nonlocal measurements are taken at a sufficiently distant from the current injection terminal, the extracted mirror Hall conductivity approached quantized theoretical prediction over a wide range of energies.

\section{Discussion and conclusion}\label{sec12}

The strong 3D TI $\mathrm{Bi}_{2}\mathrm{Te}_{3}$ has the rhombohedral
structure \citep{nakajima1963crystal}. The bulk structure are constructed
by the hexagonal monatomic crystal planes which are stacked along
c-axis in ABC order \citep{zhang2009topological}. Units of Te-Bi-Te-Bi-Te
form a quintuple layer (QL). The coupling is covalent between atomic
planes within a QL whereas weak between adjacent QLs, predominantly
of the van der Waals type. It is convenient to work in the hexagonal
basis $\mathbf{a}_{1}=a(\frac{\sqrt{3}}{2},-\frac{1}{2},0)$, $\mathbf{a}_{2}=a(0,1,0)$,
and $\mathbf{a}_{3}=c(0,0,1)$ with $a=4.38\mathring{\mathrm{A}}$
and $c=30.49\mathring{\mathrm{A}}$. The crystal structure belongs
to the space group $R\overline{3}m$ (No. 166), which has the Bi atoms
situated at $6c(0,0,\pm0.4005)$, the type-1 Te atoms (Te1) at $6c(0,0,\pm0.2097)$,
and type-2 Te atoms (Te2) at $3a(0,0,0)$ Wyckoff positions. The generators
of the space group are $\{1|0\}$, $\{1|0\}$,$\{2_{110}|0\}$, $\{-1|0\}$,
and $\{1|\frac{2}{3},\frac{1}{3},\frac{1}{3}\}$. The combination
of the inversion and the two-fold rotation symmetry gives rise a mirror
symmetry $\{m_{110}|0\}=\{-1|0\}\{2_{110}|0\}$. The three-fold rotation
symmetry about the z-axis produces two other mirror symmetries $\{m_{100}|0\}$
and $\{m_{010}|0\}$. Hence there are total three mirror planes in
$\mathrm{Bi}_{2}\mathrm{Te}_{3}$ which are perpendicular to $[110]$,
$[100]$ and $[010]$ axes, respectively. Thus the thin film perpendicular
to these three axes possess the mirror symmetry.

Another way to grow a thin film with mirror symmetry is to making
use of twin boundary in crystal \citep{authier2003international,medlin2010structure,lu2016stabilizing}.
A twin plane is planar stacking faults in a fixed crystallographic
direction (say, $[uvw]$) and usually has low formation energies.
This phenomenon is widely observed in group IV (e.g., Si) and III-V
(e.g., GaAs and InP) semiconductor. Here we are interested on the
$[uvw]$-oriented slab with a single twin boundary in the middle.
 Structurally, it can be described as the reversal of atomic
stacking sequence along the $[uvw]$ direction about the twin plane. For example, the $[001]$-oriented slab of topological insulator $\mathrm{Bi}_{2}\mathrm{Te}_{3}$
with a twin plane respects the mirror symmetry. It is known that the
interface between two TIs does not host any gapless interface states,
and the surface states only appear on the top and bottom surfaces.
The sample may also be made possibly by folding a TI thin film mechanically,
which technique was extensively used in field of 2D materials such
as twisted graphene \citep{cao2018unconventional}. Hence the slab
with a twin plane is an ideal material candidate to realize half quantum
mirror Hall effect.

If the thickness of the TI film is reduced to that the wave functions
of the top and bottom surface states have a spatial overlap, the surface
states gap out, and induces a tiny gap $E_g$ at the $\Gamma$
point \citep{Linder2009prb,lu2010massive}. A mirror Chern number
can be well defined for the two gaped surface bands, $C_{\chi}=0$
or $\chi$. The nontrivial case has the mirror Hall conductance $\sigma_{xy}^{M_z}=2\frac{e^{2}}{h}$,
that is actually the quantum spin Hall effect. Even in this gapped
case, if the chemical potential $\mu$ deviates from the energy
gap $E_g$, it is found that the mirror Hall conductance approaches
to $\frac{e^{2}}{h}$ very quickly from $0$ or $2\frac{e^{2}}{h}$,
which reflects the fact that the symmetry breaking term near the mirror
plane cooperates into the bands aways from the low-energy dispersions
of the surface states.

The topological robustness of the half-quantum mirror Hall effect in a mirror-symmetric topological insulator is safeguarded by time-reversal and mirror symmetries. The latter is a spatial symmetry and can be broken by surface roughness or the charge transfer from the substrate. Within the mirror operator's eigenbasis, symmetry-breaking manifests as an inter-sector coupling, characterized by the inter-mirror scattering length $l_{M_z}$ that gauges the extent of symmetry disruption. As highlighted in the "transport signature" section, an increase of the inter-sector symmetry-breaking term causes a reduction in $l_{M_z}$ and a corresponding invisibility of the observable transport phenomena associated with the half-quantum mirror Hall effect.

In summary, the half-quantum mirror Hall effect reveals a new type
of mirror-symmetry induced quantum anomaly in a time-reversal invariant
lattice TI film, giving rise to a topological metallic state of matter
with time-reversal invariance.

\section{Figures}\label{sec6}


\begin{figure}[H]
    \centering
    \includegraphics[width=0.9\textwidth]{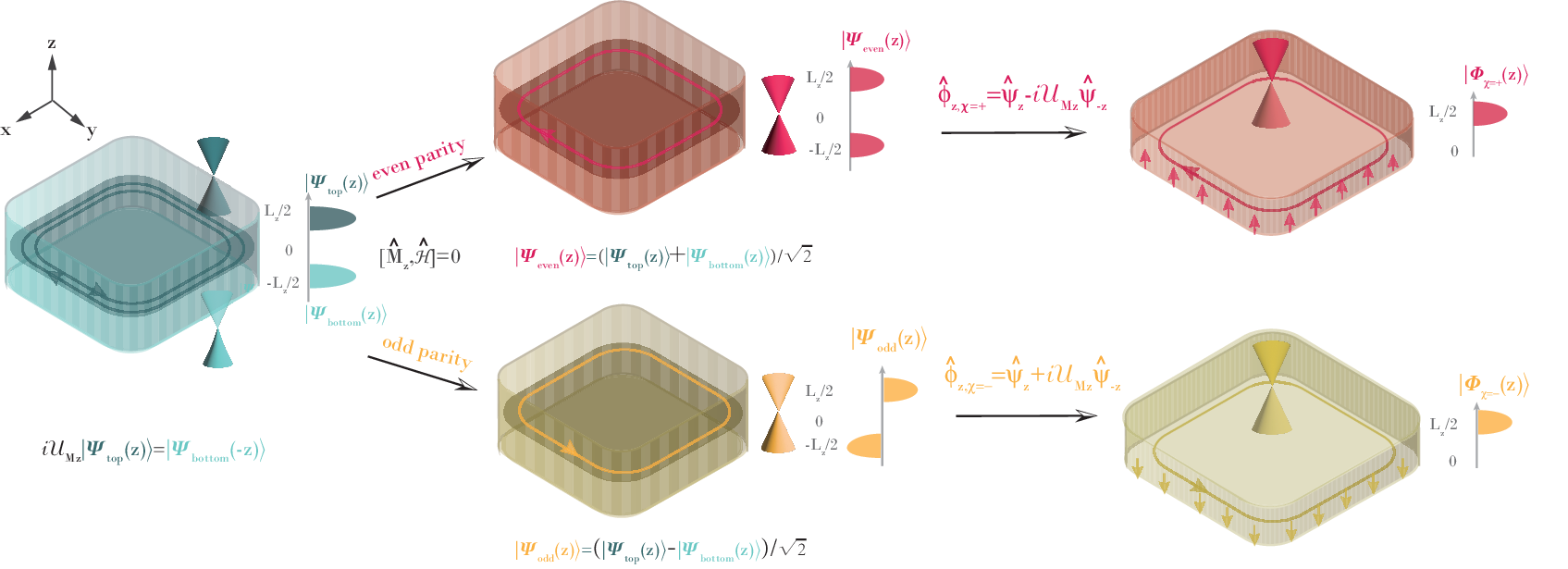}
    \caption{  Schematic of separation of a time-reversal invariant system (left panel) with
   mirror symmetry into two symmetry broken subsystems with mirror parity (middle panel) and  in the mirror operator eigenbasis its equivalence to the topological insulator thin film with a single surface subjected to a symmetry breaking term (right panel). The system characterized by definite mirror parity features a single Dirac cone within the Brillouin zone, where the wavefunction of surface states exhibits either even($\chi=1$) or odd ($\chi=-1$) parity with respect to the z-axis. After transitioning to the eigenbasis of mirror operator, the wavefunction of the surface state is distributed on a single surface within the new coordinate framework. The red and yellow arrows indicate the counter-propagating mirror current $J_{\chi=\pm}$ of the half quantum mirror Hall effect.  
}
\end{figure}

\begin{figure}[H]
    \centering
    \includegraphics[width=0.9\textwidth]{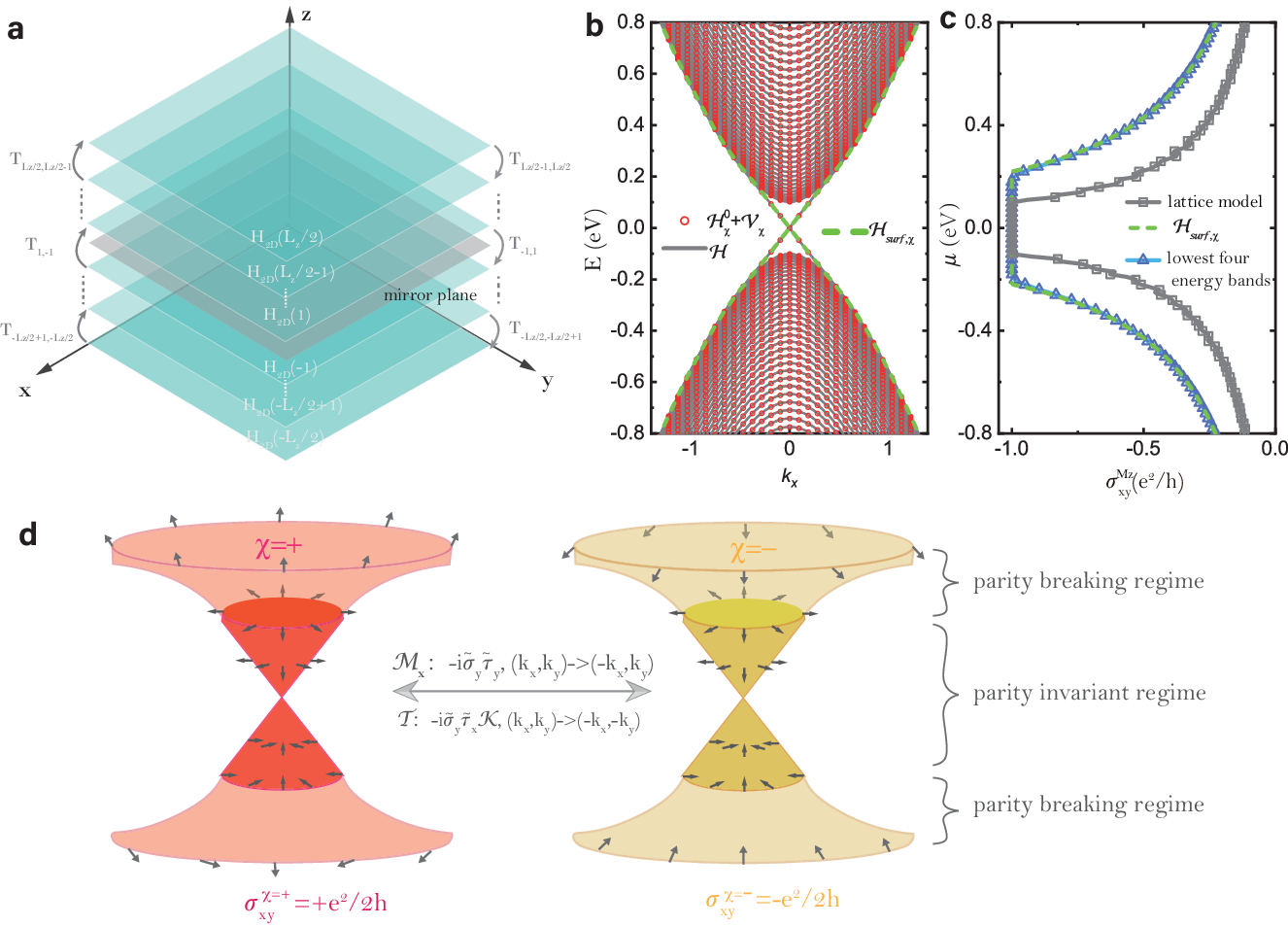}

    \caption{
$\mathbf{a.}$ Schematic diagram of a film exhibiting mirror symmetry. $H_{2D}(z)$ represents the Hamiltonian of each layer, while $T_{z,z+1}$ and  $T_{z+1,z}$ represent the hopping terms between adjacent layers in the z-direction. $\mathbf{b.}$ The band structure of topological insulator thin film (gray lines)
    based on a 3D tight-binding model calculation with open boundary conditions
    parallel to the mirror plane and periodic boundary conditions along
    the remaining two directions. The red circles represent the energy
    spectrum by diagonalizing the Hamiltonian describing half thickness
    of the film ($\hat{\mathcal{H}}_{\chi}^{0}$) with a time reversal symmetry
    breaking term $\hat{\mathcal{V}}_{\chi}$ on the bottom layer. The green dashed
    lines are from the effective Hamiltonian {[}Eq. (\ref{eq:effective_H}){]}.
    $\mathbf{c.}$ The mirror Hall conductance as a function of energy calculated
    by using Eq. (\ref{eq:mirror_hall}) based on the 3D tight-binding
    model for TI and the model for the surface states in Eq. (\ref{eq:effective_H}). The blue line marked with triangles represents the contributions from the lowest-energy four gapless bands as determined by tight-binding model.
$\mathbf{d.}$ Schematic diagram of two separated classes of the gapless Dirac cones with even and odd parity. Black arrows represent the pseudo-spin texture (i.e., $\hat{\mathbf{d}}=\langle\widetilde{\boldsymbol{\sigma}}\rangle$) of two gapless Dirac cones with opposing mirror eigenvalues $i\chi$. The parity symmetry ($\mathcal{M}_x$) is disrupted at high energy states for each miror sector because the pseudo-spin orientation is not maintained under parity operations; however, it is preserved within the low energy states.
}
\end{figure}

\begin{figure}[H]
    \centering
    \includegraphics[width=0.9\textwidth]{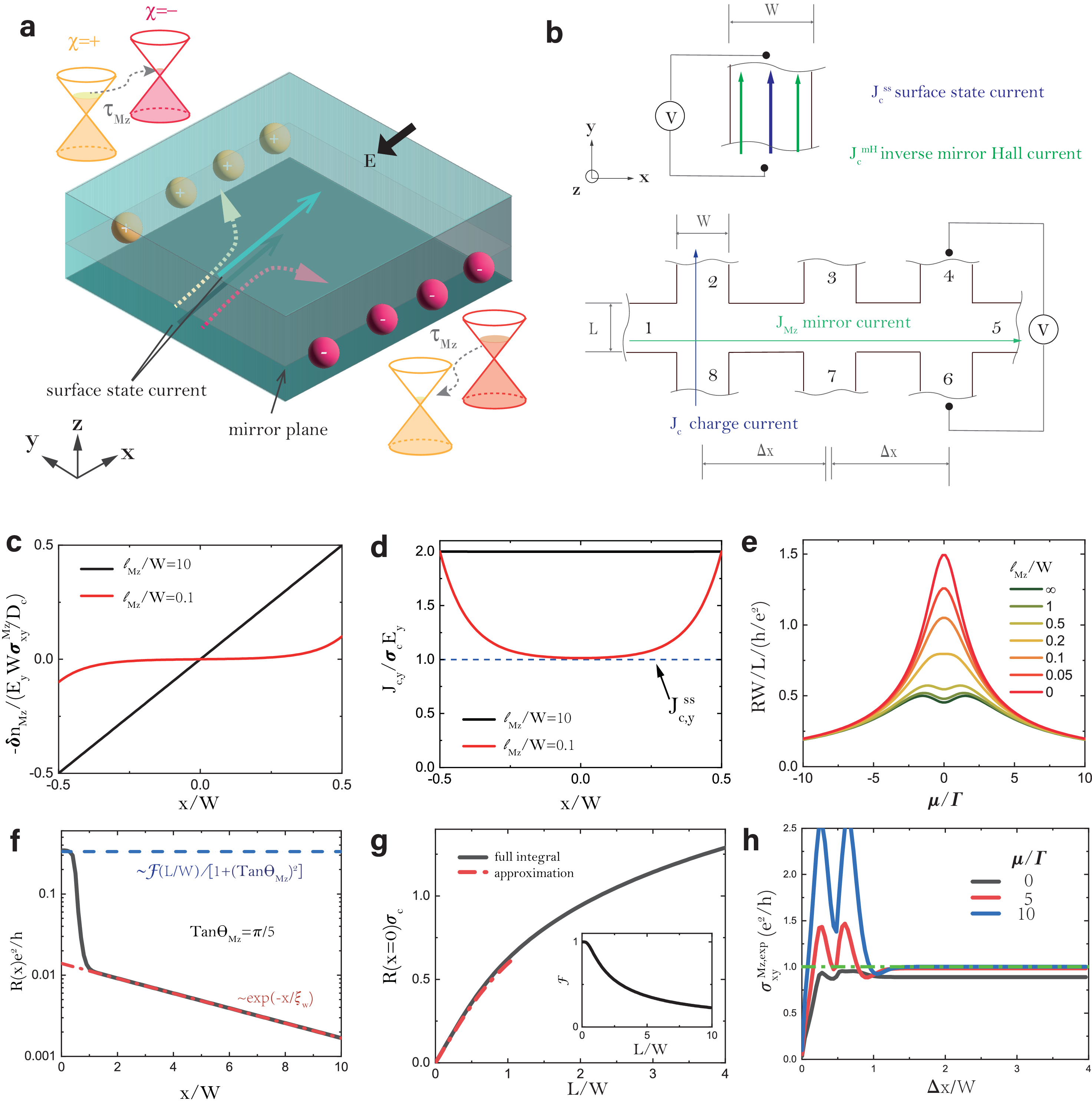}

    \caption{
$\mathbf{a.}$ Schematic of mirror Hall effect analyzed in this
    work. The electrons in $\chi=+$ and $\chi=-$ sectors are denoted by the red and yellow filled circles, respectively. The symbol "E" stands for the in plane electric field. The dark green arrows with solid lines signify the currents generated by the surface states. $\tau_{M_z}$ denotes the scattering time between two mirror sectors.
 $\mathbf{b.}$ A sketch of the two-terminal transport setup and multi-terminal transport setup.  
$\mathbf{c.}$ The spatial distributions of polarization density between
    mirror eigenvalue sectors $\delta n_{M_z}$ and $\mathbf{d.}$ the charge current
    density $J_{c,y}$ in two limiting regimes. Blue dashed line in panel
    $\mathbf{d}$ represents the current density from the conducting surface states.
  $\mathbf{e.}$ The two terminal resistance as a function of chemical potential
    $E$ for different relative ratios $l_{M_z}/W$. The longitudinal conductivity of Dirac surface states can be calculated by $\sigma_c=\frac{e^2}{\pi h}[1+(\frac{\mu}{\Gamma}+\frac{\Gamma}{\mu})\arctan(\frac{\mu}{\Gamma})]$. $\Gamma$
    is imaginary part of the self-energy around the Dirac point which
    can be determined within self-consistent Born approximation. 
$\mathbf{f.}$ The  resistance $R(x)$ as a function of the probe position $x$ according to Eq. (\ref{eq:4resistance}) (black line). The red dashed line represents the the nonlocal
resistivity obtained when the voltage measurement is taken at a location sufficiently distant from the current injection terminal, as described by Eq. (\ref{eq:RNL}). The blue dashed line corresponds to the approximation result given by Eq. (\ref{eq:R2828}), which applies when voltage measurement is conducted at the current injection terminal.
$\mathbf{g.}$ $R(x=0)$ as a function of $L/W$ from Eq. (\ref{eq:RNL})(black solid line) and Eq. (\ref{eq:R2828}) (red dashed line). The inset displays $\mathcal{F}$ versus $L/W$. Here we use $\tan\theta_{M_z}=\pi/5$.
$\mathbf{h.}$ The measured mirror Hall conductivity from Eq. (\ref{eq:exp_mirror_hall}) calculated from Eq. (\ref{eq:4resistance}) for different chemical potentials.We have adopted $W$ as the unit of length. For  Figs. $\mathbf{f}, \mathbf{g},$ and $\mathbf{h}$, we have set $l_{M_z}/W = 4$. In Figs. $\mathbf{f}$ and $\mathbf{h}$, a ratio of $L/W = 0.3$ is utilized.
}
\end{figure}

\section{Methods}\label{sec11}

\paragraph*{Tight-binding model for calculations-}

Following Hamiltonian of topological insulator thin film stated, finite along $z$-direction while homogeneous in $x-y$ plane, a concrete case is chosen as 
\begin{equation}
	\hat{\mathcal{H}} = \sum_{\boldsymbol{l}} \psi_{\boldsymbol{l}}^{\dagger} H_0 \psi_{\boldsymbol{l}} + \sum_{\boldsymbol{l},\boldsymbol{\delta}} \left( \psi^{\dagger}_{\boldsymbol{l}} T_{\boldsymbol{l},\boldsymbol{l}+\boldsymbol{\delta}} \psi_{\boldsymbol{l}+\boldsymbol{\delta}} + \mathrm{h.c.} \right),
\end{equation}
where $\boldsymbol{l}\equiv(\mathbf{r},z)=(x,y,z)$ denotes  the lattice sites in three-dimensional space and $\boldsymbol{\delta} = \{\hat{\mathbf{x}},\hat{\mathbf{y}},\hat{\mathbf{z}}\}$ denotes the unit vectors for hopping in three directions and the definition reads 
\begin{gather*}
	 H_0 = (m_0 - 2\sum_{\boldsymbol{\delta}}t_{\boldsymbol{\delta}})\sigma_0 \tau_z, ~T_{\boldsymbol{l},\boldsymbol{l}+\boldsymbol{\delta}} = t_{\boldsymbol{\delta}}\sigma_0 \tau_z - i\frac{\lambda_{\boldsymbol{\delta}}}{2}\sigma_{\boldsymbol{\delta}} \tau_x.
\end{gather*}
In Eq. (\ref{eq:Hamiltonian}), $H_{\hat{\boldsymbol{x}}}=T_{\boldsymbol{l},\boldsymbol{l}+\hat{\boldsymbol{x}}}$ and $H_{\hat{\boldsymbol{y}}}=T_{\boldsymbol{l},\boldsymbol{l}+\hat{\boldsymbol{y}}}$. By adopting periodic boundary condition in $x-y$ plane the thin film Hamiltonian is obtained as 
\begin{equation}
	\hat{\mathcal{H}} = \sum_{\mathbf{k},z} \left(\psi_{\mathbf{k},z}^{\dagger} H_{2D} \psi_{\mathbf{k},z} + \psi_{\mathbf{k},z}^{\dagger}T_{z,z+1}\psi_{\mathbf{k},z + 1} + \mathrm{h.c.}\right),
\end{equation}
with the in-plane Hamiltonian $H_{2D}=H_0+\sum_{\boldsymbol{\delta}=\hat{\boldsymbol{x}},\hat{\boldsymbol{y}}}(T_{\boldsymbol{l},\boldsymbol{l}+\boldsymbol{\delta}}e^{i\boldsymbol{\delta}\cdot\mathbf{k}}+T_{\boldsymbol{l}+\boldsymbol{\delta},\boldsymbol{l}}e^{-i\boldsymbol{\delta}\cdot\mathbf{k}})$ which can be calculated as
\begin{gather*}
    H_{2D} = \lambda_{\parallel} [\sin (k_x a) \sigma_x  \tau_x + \sin(k_y b )\sigma_y  \tau_x] + M_0 (\mathbf{k}) \sigma_0  \tau_z,\\
\end{gather*}
where $ M_0 (\mathbf{k}) = m_0(\mathbf{k}) - 2t_{\perp}$, and
\begin{equation*}
    m_0(\mathbf{k}) = m_0 - 4t_{\parallel} \left(\sin^2\frac{k_x a}{2} + \sin^2 \frac{k_y b}{2}\right).
\end{equation*}
Notice that we have adopted a homogeneous setup in $x-y$ plane $t_{\boldsymbol{\hat{x}}}=t_{\boldsymbol{\hat{y}}}=t_{\parallel}$, $\lambda_{\boldsymbol{\hat{x}}}=\lambda_{\boldsymbol{\hat{y}}}=\lambda_{\parallel}$, $t_{\boldsymbol{\hat{z}}}=t_{\perp}$, and $\lambda_{\boldsymbol{\hat{z}}}=\lambda_{\perp}$. Nevertheless, to illustrate the nature, 
we choose a simple set of parameters as $\lambda_{\parallel} = \lambda_{\perp} = t_{\parallel} = t_{\perp} = 0.5 \mathrm{eV},~m_0 = 0.1 \mathrm{eV}$,
together with unit lattice parameters as $a = b = c = 1\mathrm{nm}$ in the whole numerical calculation. The thickness of the film is set as $L_z=80$.

\paragraph*{Derivation of Gapless Dirac Cone Model with Parity Symmetry Breaking}

We solve the Hamiltonian $\hat{\mathcal{H}}=\hat{\mathcal{H}}_{1d}+\hat{\mathcal{H}}_{\parallel}$ by separating it into two parts: the parallel part is given by $\hat{\mathcal{H}}_{\parallel}=\sum_{\mathbf{k}}\sum_{z}  \hat{\psi}_{\mathbf{k},z}^{\dagger}\lambda_{\parallel} [\sin (k_x a) \sigma_x  \tau_x + \sin(k_y b )\sigma_y  \tau_x]\hat{\psi}_{\mathbf{k},z}$,
and  the one-dimensional (1-D) Hamiltonian part $\hat{\mathcal{H}}_{1d}$, which is described by
\begin{equation}
	\hat{\mathcal{H}}_{1d} = \sum_{\mathbf{k}}\sum_{z} \left( \hat{\psi}_{\mathbf{k},z}^{\dagger} M_0 (\mathbf{k}) \sigma_0  \tau_z \hat{\psi}_{\mathbf{k},z} + \hat{\psi}^{\dagger}_{\mathbf{k},z} T_{z,z+1} \hat{\psi}_{\mathbf{k},z+1} + \mathrm{h.c.} \right).
\end{equation} 
The eigenvalue problem of the 1-D Hamiltonian with respect to boundary condition is 
\begin{equation}
	\begin{cases}
		H_{1d}(\mathbf{k}) \ket{\Phi(\mathbf{k})} = E(\mathbf{k}) \ket{\Phi(\mathbf{k})},\\
		\Phi(\mathbf{k},\pm l/2) = 0,
	\end{cases}
\end{equation}
where $l = L_z + 1$ with $L_z$ total sites number along $z$. $H_{1d}(\mathbf{k})$ takes the form of a  block tridiagonal matrix in terms of the layer index with  dimension as $4L_z\times4L_z$, defined as $\hat{\mathcal{H}}_{1d}=\sum_{\mathbf{k}}\hat{\Psi}_{\mathbf{k}}^{\dagger}H_{1d}(\mathbf{k})\hat{\Psi}_{\mathbf{k}}$, where $\hat{\Psi}_{\mathbf{k}}$ represents the collective spinor encompassing all the layer components. Solving the set of equations above gives eigenstates read 
\begin{equation}
	\begin{aligned}
		\Phi_1(l_z) &= \begin{pmatrix}
			\varphi(+,l_z) \\ 0
		\end{pmatrix},\ \Phi_2(l_z) = \begin{pmatrix}
			0 \\ \chi(-,l_z)
		\end{pmatrix}, \\
		\Phi_3(l_z) &= \begin{pmatrix}
			\chi(+,l_z) \\ 0
		\end{pmatrix},\ \Phi_4(l_z) =\begin{pmatrix}
			0 \\ \varphi(-,l_z)
		\end{pmatrix},
	\end{aligned}
\end{equation}
where 
\begin{equation}
	\begin{cases}
		\varphi(s,l_z) = C \begin{pmatrix}
			-is\lambda_{\perp} f_+(l_z) \\ t_{\perp} \eta f_-(l_z)
		\end{pmatrix},\ E \\ 
		\chi(s,l_z) = C \begin{pmatrix}
			t_{\perp} \eta f_-(l_z) \\ is\lambda_{\perp} f_+(l_z)
		\end{pmatrix},\ -E
	\end{cases},
\end{equation}
with $C$ the norm, and $\pm E$ refer to corresponding eigenvalue which could be solved consistently in a closed manner with equations
\begin{equation}
	\begin{cases}
		E = M + 2 t_{\perp} \frac{\cos \xi_1 g(\xi_1) - \cos \xi_2 g(\xi_2)}{g(\xi_1) - g(\xi_2)},\\
		\cos \xi_{\alpha} = \frac{-M t_{\perp} + ( - 1)^{\alpha - 1}\sqrt{M^2 t_{\perp}^2 - (t_{\perp}^2 - \lambda_{\perp}^2/4)(M^2 + \lambda_{\perp}^2 - E^2)}}{2(t_{\perp}^2 - \lambda_{\perp}^2/4)},
	\end{cases}
\end{equation}
where $\alpha = 1,2$ and $M$ is referred to $M_0(\mathbf{k})$. For definitions of $f_{\pm}$ and $\eta$, please refer to the supplementary material. 

Our solution is exact with $\mathbf{k}$-dependence and includes all solutions $E_n(\mathbf{k}),~n = 1,\cdots,L_z$, with $n = 1$ denoting possible non-trivial zero-mode state. 
Then by projecting the original Hamiltonian onto the obtained $4 L_z$ eigenstates, we get 
the effective Hamiltonian 
\begin{equation}
    \begin{aligned}
        H_{\text{EFF}}(\mathbf{k}) &= \bigoplus_{n = 1}^{L_z} \left[\lambda_{\parallel}  (\sin (k_x a)\widetilde{\sigma}_x + \sin (k_y b)\widetilde{\sigma}_y)\widetilde{\tau}_0 +  E_n  \widetilde{\sigma}_z\widetilde{\tau}_z\right] = \bigoplus_{n , \chi } h_{n,\chi}(\mathbf{k}),
    \end{aligned}
\end{equation}
where $\chi = \pm$ is the mirror label, and due to the explicit direct sum form, we separate $n = 1$ block and make equivalence with 
\begin{equation}
	H_{\mathrm{surf},\chi} = h_{1,\chi} = \lambda_{\parallel} (\sin (k_x a)\widetilde{\sigma}_x + \sin (k_y b)\widetilde{\sigma}_y) +  \chi \Delta(\mathbf{k}) \widetilde{\sigma}_z,~\Delta(\mathbf{k}) \equiv E_1(\mathbf{k}),
\end{equation} 
which is Eq. (2) in the main text. Meanwhile, it could be proved (please refer to the supplementary material) that in the thick limit, 
\begin{equation}
	\Delta(\mathbf{k}) \simeq \Theta(-m_0(\mathbf{k})) m_0(\mathbf{k}).
\end{equation}

\paragraph*{Topological field theory for quantum mirror Hall effect on a lattice}

We start with the four-band Hamiltonian of 3D TI thin films to illustrate
the topological field theory for quantum mirror Hall effect, which
can be written as $H_{0}(\mathbf{k})=\lambda_{\parallel}(\sin k_{x}\alpha_1+\sin k_{y}\alpha_2)+\Delta(\mathbf{k})\alpha_{3}$
with the Dirac matrices $\alpha_{1,2}=\tau_{0}\sigma_{1,2}$, $\alpha_{3}=\tau_{3}\sigma_{3}$.
It is should be emphasized that the four-band theory is constructed
on lattice with finite 2D Brillouin zone. The time-ordered Green function
is $\mathcal{G}_{0}(k)=[\omega-\mathbf{d}\cdot\overrightarrow{\alpha}(1-i\eta)]^{-1}$where
$k^{\mu}=(\omega,\mathbf{k})_{\mu}$, $\mathbf{d}(\mathbf{k})=(\lambda_{\text{\ensuremath{\parallel}}}\sin k_{x},\lambda_{\text{\ensuremath{\parallel}}}\sin k_{y},\Delta(\mathbf{k}))$
and $\eta$ is infinitsimal small quantity. We are interested in the
systems with a horizontal mirror symmetry $M_{z}=\alpha_{1}\alpha_{2}\alpha_{3}$
which maps two surface states into each other. In order to study a
linear electromagnetic response in the thin film system, we include
the electromagnetic fields $\mathcal{A}$ and the mirror electromagnetic
fields $\mathcal{B}$ which are coupled to the charge current and
mirror current respectively through the interaction term $H_{\mathrm{gauge}}=j_{c,\mu}\mathcal{A}^{\mu}+j_{M_z,\mu}\mathcal{B}^{\mu}$,
where the electric current density operator in the momentum space
is given by $j_{c,\mu}=\partial_{k^{\mu}}\mathcal{G}_{0}^{-1}(k)$ and $j_{M_z,\mu}=iM_{z}j_{c,\mu}$ denotes the mirror current.
By integrating out the fermions in the action,
the effective action for gauge fields $S_{\mathrm{eff}}[\mathcal{A},\mathcal{B}]$
can be obtained by expanding to the quadratic order\citep{Lapa2019prb}
\begin{align}
\mathcal{S}_{\mathrm{eff}} & =\frac{1}{2}\int_{q}\left(\mathcal{A}^{\mu}(-q)\Pi_{\mu\nu}^{cc}(q)\mathcal{A}^{\nu}(q)+2\mathcal{A}^{\mu}(-q)\Pi_{\mu\nu}^{cM_z}(q)\mathcal{B}^{\nu}(q)+\mathcal{B}^{\mu}(-q)\Pi_{\mu\nu}^{M_zM_z}(q)\mathcal{B}^{\nu}(q)\right).\label{eq:eff_S}
\end{align}
where $\int_{q}=\int\frac{d\omega}{2\pi}\int_{BZ}\frac{d^{2}\mathbf{q}}{(2\pi)^{2}}$
and the momentum $\mathbf{q}$ integral is performed over the
whole 2D Brillouin zone. $S_{\mathrm{eff}}[\mathcal{A},\mathcal{B}]$ contains three types
of contributions: $\mathcal{A}^{\mu}\Pi_{\mu\nu}^{cc}\mathcal{A}^{\nu},2\mathcal{A}^{\mu}\Pi_{\mu\nu}^{cM_z}\mathcal{B}^{\nu}$,
and $\mathcal{B}^{\mu}\Pi_{\mu\nu}^{M_zM_z}\mathcal{B}^{\nu}$ where $\mu,\nu$
run over the space-time indices $(0,1,2)$ with the vacuum polarization
operator  as  $i\Pi_{\mu\nu}^{\kappa\kappa^{\prime}}(q)=\int_{k}\mathrm{Tr}[j_{\kappa,\mu}(k)\mathcal{G}_{0}(k+q/2)j_{\kappa^\prime,\nu}(k)\mathcal{G}_{0}(k-q/2)]$ with $\kappa,\kappa^{\prime}=c,M_z$. 
$\Pi_{\mu\nu}^{cc}$ and $\Pi_{\mu\nu}^{M_zM_z}$ are the polarization tensors
with same type photon-current vertices (charge-charge (cc) or mirror-mirror ($M_zM_z$)),
and $\Pi_{\mu\nu}^{cM_z}$ are the polarization tensors with charge-mirror (c$M_z$)
type vertices. The antisymmetric terms in $\Pi_{\mu\nu}^{cc},\Pi_{\mu\nu}^{M_zM_z}$
exactly vanishes due to the time reversal symmetry. There is no divergence
in $\Pi_{\mu\nu}^{cc}$ and $\Pi_{\mu\nu}^{M_zM_z}$ as the momentum integral
is performed over a finite Brillouin zone due to the lattice regularization.
The quantity $\Pi_{\mu\nu}^{cM_z}(q)$ only contains the antisymmetric
terms $\Pi_{\mu\nu}^{cM_z}(q)=\frac{C_{M_z}}{\pi}\epsilon_{\mu\nu\zeta}q_{\zeta}$ and the mutual Chern-Simons theory for $\mathcal{A}_{\mu}$
and $\mathcal{B}_{\nu}$ can be evaluated as 
\begin{align*}
S_{\mathrm{eff}} & [\mathcal{A},\mathcal{B}]=\frac{C_{M_z}}{\pi}\int d^{3}x\epsilon^{\mu\nu\varsigma}\mathcal{A}_{\mu}\partial_{\varsigma}\mathcal{B}_{\nu},\quad \mathrm{with} \quad C_{M_z}=\int_{BZ}\frac{d^{2}\mathbf{k}}{4\pi}\hat{\mathbf{d}}\cdot\partial_{k_{x}}\hat{\mathbf{d}}\times\partial_{k_{y}}\hat{\mathbf{d}}.
\end{align*}
where $\epsilon^{\mu\nu\varsigma}$ is Levi-Civita symbol and $\hat{\mathbf{d}}=\mathbf{d}/|\mathbf{d}|$.
 For the four-band lattice Hamiltonian $H_{0}(\mathbf{k})$,
we have $C_{M_z}=\frac{\mathrm{sgn(\Delta(\pi,\pi))}}{2}$ which is a half-integer
with its sign determined by the sign of $\Delta(\pi,\pi)$. The crossing
Chern-Simons term corresponds a half quantum mirror Hall effect $\langle j_{M_z}^{\nu}\rangle=\frac{\delta S_{\mathrm{eff}}}{\delta\mathcal{B}_{\nu}}=-\frac{\mathrm{sgn}(\Delta(\pi,\pi))}{2\pi}\epsilon^{\mu\nu\varsigma}\partial_{\varsigma}\mathcal{A}_{\mu}$.

If we now focus on the low-energy effective model of the lattice four-band
Hamiltonian by neglecting higher energy states $(\propto\Delta(\mathbf{k}))$,
which can be expressed as $H_{0}^{low}(\mathbf{k})=\lambda_{\parallel}(k_{x}\alpha_{1}+k_{y}\alpha_{2})$.
There is a linear ultraviolet divergence in $\Pi_{\mu\nu}^{cc}(q)$ and
$\Pi_{\mu\nu}^{M_zM_z}(q)$ which should be regularized by Pauli-Villars
method in a gauge-invariant way\citep{Redlich1984prl}. In the Pauli-Villars regularization
approach, we need to introduce a second Dirac field mass $M\alpha_{3}$.
In the limit ($M\to\infty$), the regulator field decouples from the
theory, which removes the divergence in $\Pi_{\mu\nu}^{cc}$ and $\Pi_{\mu\nu}^{M_zM_z}$,
leaving a finite contribution for the crossed polarization tensor
$\Pi_{\mu\nu}^{cM_z}(q)=\frac{\mathrm{sgn}(M)}{2\pi}\epsilon_{\mu\nu\zeta}q_{\zeta}$.
This also induces a crossing Chern-Simons term and corresponds to
a half-quantum mirror Hall effect.We demonstrate that for a time reversal symmetric TI thin film with time reversal symmetry and mirror symmetry, a new quantum anomaly exists, which manifests itself as a half-quantum mirror Hall effect.

\paragraph*{Charge transport associated with the half quantum mirror Hall effect}

In this section, we present the macroscopic theory for the transport
properties associated with the half quantum mirror Hall effect. The
generalized Ohm's law for currents from two Mirror eigenstates:
\[
J_{\chi,i}(\mathbf{r})=-\sigma_{ij}^{\chi}\partial_{j}\phi(\mathbf{r})+eD_{ij}^{\chi}\partial_{j}\delta n_{\chi}(\mathbf{r})
\]
where $\phi(\mathbf{r})$ is the electric potential, $i,j=x,y$ are
spatial coordinates, $\delta n_{\chi}(\mathbf{r})$ is the variation
of the charge density in mirror sector $\chi$ due to the transport,
$\sigma_{ij}^{\chi}$ is the homogeneous conductivity tensor, and
$D_{ij}^{\chi}$ is the corresponding diffusion coefficient tensor.
A repeated spatial index obeys the Einstein's summation convention.
The first and second terms on the right-hand side are the drift current
due to the electric field and the diffusion current due to the inhomogeneity
of the electron density. The diffusion constant $D_{ij}^{\chi}$ is
given by the Einstein relation $e^{2}D_{ij}^{\chi}=\sigma_{ij}^{\chi}S^{\chi}$
where $S^{\chi}=\partial\mu_{\chi}/\partial n_{\chi}=\nu_{\chi}^{-1}$
is the static stiffness and $\nu_{\chi}$ is the thermodynamics density
of state or the compressibility. We have neglected the inter-mirror
interaction such that the inter-mirror elements of the conductivity
and stiffness matrices vanish. We can define the total charge density
(c) $\delta n_{c}(\mathbf{r})=\sum_{\chi}\delta n_{\chi}(\mathbf{r})$
and the mirror polarization density ($M_{z}$) $\delta n_{M_z}(\mathbf{r})=\sum_{\chi}\chi\delta n_{\chi}(\mathbf{r})$.
Similarly, the total charge current density is determined by $J_{c,i}=\sum_{\chi}J_{\chi,i}$
and the total mirror current density is $J_{M_{z},i}=\sum_{\chi}\chi J_{\chi,i}$
which can be obtained as 
\begin{align}
J_{c,i} & =\delta_{ij}\left(-\sigma_{c}\partial_{j}\phi+eD_{c}\partial_{j}\delta n_{c}\right)+eD_{H}^{M_z}\epsilon_{ij}\partial_{j}\delta n_{M_z},\nonumber \\
J_{M_z,i} & =\delta_{ij}D_{c}\partial_{j}\delta n_{M_{z}}+\epsilon_{ij}\left(-\sigma_{H}^{M_z}\partial_{j}\phi+eD_{H}^{M_z}\partial_{j}\delta n_{c}\right),\label{eq:current_density}
\end{align}
where we have introduced the longitudinal charge conductivity $\sum_{\chi}\sigma_{ij}^{\chi}=\delta_{ij}\sigma_{c}$,
the mirror Hall conductivity $\sum_{\chi}\chi\sigma_{ij}^{\chi}=\epsilon_{ij}\sigma_{H}^{M_z}$,
longitudinal charge diffusion constant $\frac{1}{2}\sum_{\chi}D_{ij}^{\chi}=\delta_{ij}D_{c}$,
and the mirror Hall diffusion constant $\frac{1}{2}\sum_{\chi}\chi D_{ij}^{\chi}=\epsilon_{ij}D_{H}^{M_z}$
with $\delta_{ij}$ and $\epsilon_{ij}$ as the Kronecker delta and
Levi-Civita symbols. Equations (\ref{eq:current_density}) establish
a linear relationship between the densities and currents in the presence
of the electric potential. To solve these equations, we also require
the two continuity equations :
\begin{align*}
\partial_{i}J_{c,i} & =0,\\
\partial_{i}J_{M_{z},i} & =e\delta n_{M_{z}}/\tau_{M_{z}}
\end{align*}
where $\tau_{M_{z}}$ is a phenomenological relaxation time due to
the inter-mirror scattering which equilibrates the two mirrors sectors
relaxing the system to a steady state. By combining the continuity
equation with Eqs. (\ref{eq:current_density}), we can obtain the
electric field inside the system obeys the Laplace equation $\nabla^{2}\phi=0$
and a diffusion equation for the mirror polarization density 
\begin{equation}
D_{c}\boldsymbol{\nabla}^{2}\delta n_{M_{z}}=\frac{\delta n_{M_{z}}}{\tau_{M_{z}}}-\boldsymbol{\nabla}\times\left(-\frac{\sigma_{H}^{M_{z}}}{e}\boldsymbol{\nabla}\phi+D_{H}^{M_{z}}\boldsymbol{\nabla}\delta n_{c}\right)\cdot\hat{z}.\label{eq:diffusion_equation}
\end{equation}
We use the local charge neutrality constraint $\delta n_{c}\approx0$
and $\boldsymbol{\nabla}\times\left(\sigma_{H}^{M_{z}}\boldsymbol{\nabla}\phi\right)=\sigma_{H}^{M_{z}}E_{y}[\delta(x-W/2)-\delta(x+W/2)]$
takes a delta-function value at the boundary between topologically
non-trivial and trivial (vacuum) regions where $E_{i}=-\partial_{i}\phi$
is the electric field. The general solution for the diffusion equation
is 
\[
\delta n_{M_{z}}(\mathbf{r})=Ae^{-x/l_{M_{z}}}+Be^{x/l_{M_{z}}}
\]
 with $l_{M_{z}}=\sqrt{D_{c}\tau_{M_{z}}}$ the mirror diffusion length.
This equation needs to be supplemented by suitable boundary conditions.
Here, we consider a two-terminal transport measurement and a charge
current is driven between injector and collector electrodes by electric
field. In this situation, the boundary condition is on the mirror
current density $J_{M_{z},x}(x=\pm W/2,y)=0$ which implies that no
mirror current can flow outside the sample. By combining with Eqs.
(\ref{eq:current_density}), we can obtain the boundary constraint
on the mirror polarization density 
\[
\left(\partial_{x}\delta n_{M_{z}}-\frac{\tan\theta_{M_{z}}}{1+\tan^{2}\theta_{M_{z}}}\frac{J_{c,y}}{D_{c}}\right)\Big|_{x=\pm W/2}=0
\]
 where $\theta_{M_{z}}=\tan^{-1}(\sigma_{H}^{M_{z}}/\sigma_{c})$
is the mirror Hall angle. By solving the differential equations with
the boundary conditions, we arrive the the polarization density in Eq. (\ref{eq:charge_polarization}) and
the charge current density in Eq. (\ref{eq:2terminal_current_density})

For a multi-terminal measurement illustrated in the lower panel of Fig. 3b, the current $I$ is applied
along the y-direction, flowing from from terminal 8 to terminal 2. We impose periodic boundary conditions
in the $x$-direction.
Then the problem can be solved by Fourier transforming all the physical
quantities in the $x$-direction $\widetilde{f}(k,y)=\int_{-\infty}^{\infty}dxe^{-ikx}f(x,y)$.
Within a conductor at electrostatic equilibrium, the electric potential $\phi$ satisfies the Laplace equation $\nabla^{2}\phi=0$. Furthermore, in accordance with the diffusion equation [Eq. (\ref{eq:diffusion_equation})],
we can assume the solutions of the form $\widetilde{\phi}(k,y)=a\cosh(ky)+b\sinh(ky)$ for the potential,
and $\delta\widetilde{n}_{M_{z}}(k,y)=c\cosh(\omega(k)y)+d\sinh(\omega(k)y)$ for the mirror polarization density
where $\omega(k)=\sqrt{k^{2}+l_{M_{z}}^{-2}}$. By using the the boundary
conditions for the charge current density $J_{c,y}(x,y=\pm \frac{L}{2})=\frac{I}{W}\Theta(\frac{W}{2}-|x|)$
and mirror current density $J_{M_{z},y}(x,y=\pm \frac{L}{2})=0$, we can obtain
the resistance as a function of the probe position $x$ according
to the fundamental definition $R(x)=[\phi(x,\frac{L}{2})-\phi(x,-\frac{L}{2})]/I$
\begin{equation}
R(x)  =\frac{2}{\sigma_{c}}\int\frac{dk}{2\pi}\frac{e^{ikx}\sin(\frac{kW}{2})/(kW/2)\omega(k)}{\omega(k)k\coth\left(\frac{Lk}{2}\right)+k^2\tan^{2}(\theta_{M_{z}})\coth\left(\frac{\omega(k)L}{2}\right)}.\label{eq:4resistance}
\end{equation}
In two limiting regimes, we derive the analytical
expressions for nonlocal resistivity in Eq. (\ref{eq:RNL}) and  for the  resistivity  as given in Eq. (\ref{eq:R2828}) when current injection and voltage measurement are conducted  on the same terminals.

\backmatter

\bmhead{Acknowledgments}

This work was supported by the National Key R\&D Program of China
under Grant No. 2019YFA0308603 and the Research Grants Council, University
Grants Committee, Hong Kong under Grant Nos. C7012-21G and 17301823.

\bmhead{Author contributions}

S.-Q. S conceived the project. B. F. and K.-Z.B. performed the theoretical analysis and simulation. B. F. and S.-Q. S. wrote the manuscript with inputs from all authors. All authors contributed to the discussion of the results. 

\bmhead{Competing interests}

The authors declare no competing interests.

\bmhead{Additional information}

$\mathbf{Supplementary}$ $\mathbf{information}$ The online version contains
supplementary material available at XXX.

$\mathbf{Correspondence}$ and requests for materials should be addressed to Shun-Qing Shen.

\begin{appendices}




\end{appendices}


\bibliography{sn-bibliography}

\end{document}